\documentclass[12pt,a4paper]{article}

\usepackage{latexsym,amsfonts,amssymb,graphicx}

\newcommand{\be}{\begin{equation}}
\newcommand{\ee}{\end{equation}}
\newcommand{\bea}{\begin{eqnarray}}
\newcommand{\eea}{\end{eqnarray}}
\newcommand{\nn}{\nonumber\\}

\renewcommand{\theequation}{\arabic{section}.\arabic{equation}}

\def\H{{\cal H}}                                 %
\def\cR{{\cal R}}                                %
\def\cA{{\cal A}}                                %
\def\cB{{\cal B}}                                %
\def\ri{{\mathrm i}}                             %
\def\pa{{\partial}}                              %
\def\cD{{\cal D}}                                %
\def\cH{{\cal H}}                                %
\def\cS{{\cal S}}                                %
\def\cT{{\cal T}}                                %
\def\bR{{\bf R}}                                 %
\def\bZ{{\bf Z}}                                 %
\def\cC{{\cal P}} 
\def\x{{\theta}}  

\begin{document}

\vspace*{0.5cm}
\begin{center}
{\Large \bf  Inequivalent quantizations of the three-particle
Calogero model constructed by separation of variables}
\end{center}

\vspace{0.5cm}

\begin{center}
L. Feh\'er${}^{a,}$\footnote{Postal address: MTA KFKI RMKI,
1525 Budapest 114, P.O.B. 49,  Hungary.
e-mail: lfeher@rmki.kfki.hu},
I. Tsutsui${}^{b}$\ and T. F\"ul\"op${}^{b}$\  \\

\bigskip

{\em ${}^a$Department of Theoretical Physics\\
MTA  KFKI RMKI and University of Szeged, Hungary}

\bigskip
{\em ${}^b$Institute of Particle and Nuclear Studies\\
High Energy Accelerator Research Organization (KEK)\\
Tsukuba 305-0801, Japan}
\end{center}

\vspace{0.5cm}

\begin{abstract}

We quantize the 1-dimensional 3-body problem with harmonic and
inverse square pair potential by separating the Schr\"odinger equation
following the classic work of Calogero, but allowing all
possible self-adjoint
boundary conditions for the angular and radial Hamiltonians.
The inverse square coupling constant is taken to be $g=2\nu (\nu-1)$
with $\frac{1}{2} <\nu< \frac{3}{2}$ and then the angular Hamiltonian
is shown to admit a 2-parameter family of inequivalent quantizations
compatible with the dihedral $D_6$ symmetry of its potential term
$9 \nu (\nu -1)/\sin^2 3\phi$.
These are parametrized by a matrix $U\in U(2)$ satisfying
$\sigma_1 U \sigma_1 = U$, and in all cases we describe the
qualitative features of the angular eigenvalues and classify
the eigenstates under the $D_6$ symmetry and its
$S_3$ subgroup generated by the particle exchanges.
The angular eigenvalue $\lambda$ enters
the radial Hamiltonian through the potential
$(\lambda -\frac{1}{4})/r^2$ allowing a
1-parameter family of self-adjoint boundary conditions
at $r=0$ if $\lambda <1$.
For $0<\lambda<1$ our analysis of the radial Schr\"odinger
equation is consistent with previous results on the possible
energy spectra, while for
$\lambda <0$
it shows that the energy is not bounded from below
rejecting those $U$'s admitting such eigenvalues as physically
impermissible.
The permissible self-adjoint angular Hamiltonians include, for example,
the cases $U=\pm {\bf 1}_2, \pm \sigma_1$, which are explicitly solvable
and are presented in detail.
The choice $U=-{\bf 1}_2$ reproduces Calogero's quantization, while
for the choice  $U=\sigma_1$ the system is smoothly connected to
the harmonic oscillator in the limit $\nu \to 1$.

\end{abstract}

\bigskip

\newpage

\section{Introduction}
 \setcounter{equation}{0}

 The Calogero model \cite{Cal3,CalN} of $N$ identical particles on the
line subject to combined  inverse square and
  harmonic interaction potential is  extremely popular because of its
 exact solvability  and its connections  to many interesting problems
in physics and mathematics.
(See, for instance, \cite{Calbook} and references therein.)
 The Hamiltonian of the system is formally given by
 \begin{equation}
H= - \frac{\hbar^2}{2m} \sum_{i=1}^N \frac{\pa^2}{\pa x_i^2}
+\sum_{i=2}^N \sum_{j=1}^{i-1}
\bigl\{ \frac{1}{4} m \omega^2 (x_i - x_j)^2 + g(x_i - x_j)^{-2}  \bigr\}.
\label{1.1}\end{equation}
After separation of the centre of mass,
the $N=2$ case reduces to the study of the
1-dimensional Schr\"odinger operator
 \begin{equation}
H_y = - \frac{\hbar^2}{2m}  \frac{d^2}{d y^2}
+  \frac{1}{2} m \omega^2 y^2 + \frac{g}{2} y^{-2}.
\label{1.2}\end{equation}
As is  widely known \cite{Landau},
the spectrum of $H_y$  cannot be bounded from below if
$g< - \frac{\hbar^2}{4m} $,
and for this reason\footnote{The energy can still be bounded from
below if $g= -\frac{\hbar^2}{4m}$,
but  this case  would  require a separate treatment.}
Calogero assumed in his work that
$g> - \frac{\hbar^2}{4m}$.
For selecting the  `admissible wave functions'  he imposed the criterion
that the associated probability current should vanish at the
locations where any two particles collide.
This is intuitively reasonable if the inverse square potential is repulsive.

Mathematically speaking,  the selection of admissible wave functions is
equivalent to choosing  a domain on which the Hamiltonian is self-adjoint.
Concerning the 1-dimensional Hamiltonian $H_y$,
it is known (see, e.g., \cite{MT,FTC} or the books \cite{DS,Richt})
that the choice of its self-adjoint domain
is essentially unique
if $g\geq  \frac{3 \hbar^2}{4m} $,
 but there exits a family of different
possibilities parametrized by a $2\times2$ unitary matrix if
$g<\frac{3 \hbar^2}{4m} $.
In the corresponding quantizations of the $N=2$ Calogero model
the probability current does not in general vanish at the coincidence
of the coordinates of the particles.
Heuristically speaking,
if $ 0<g< \frac{3 \hbar^2}{4m} $, then
the particles can pass through each other by a tunneling effect
despite the infinitely high repulsive potential barrier.
Since this phenomenon refers to the interaction of any pairs of  particles,
one may expect it to occur also in the $N$ particle Calogero model.

The purpose of the present paper is to explore the inequivalent
quantizations of the   Calogero model under the assumption
\begin{equation}
 - \frac{\hbar^2}{4m} <  g < \frac{3 \hbar^2}{4m}\qquad\qquad (g\neq 0)
\label{1.3}\end{equation}
 focusing on the simplest non-trivial case of three particles.
We shall use separation of variables to define the quantizations.
To explain the main point,
recall that the Hamiltonian (\ref{1.1})
can be written as $H= H_0 + H_{rel}$,  where $H_0$
belongs to the
centre of mass and
\begin{equation}
H_{rel}= H_{r} + r^{-2} H_{\Omega}
\label{1.4}\end{equation}
describes the relative motion.
The relative Hamiltonian $H_{rel}$ consists of the radial operator
 \begin{equation}
H_r = - \frac{\hbar^2}{2m} \frac{d^2}{d r^2}
- \frac{\hbar^2}{2m} \frac{N-2}{r}\frac{d}{d r}
+ \frac{1}{4} N m \omega^2 r^2
\label{1.5}\end{equation}
and the angular Hamiltonian
 \begin{equation}
H_\Omega= -\frac{\hbar^2}{2m} \Delta_\Omega + g
\sum_{i=2}^N \sum_{j=1}^{i-1}  [r/(x_i - x_j)]^{2},
\label{1.6}\end{equation}
where  $r$ is  the radial coordinate on ${\bR}^{N-1}$ spanned
by the relative (Jacobi) coordinates of the particles
and $\Delta_\Omega$ is the standard Laplacian on $S^{N-2}$.
In effect, Calogero's solution amounts to constructing an orthogonal  basis of
the Hilbert space $L^2(\bR^{N-1})$ in the factorized form
$R_{E,\lambda}(r) \eta_\lambda(\Omega)$, where
$\Omega$ denotes the angle coordinates on the sphere
$ S^{N-2}\subset \bR^{N-1}$ and
\begin{equation}
H_\Omega \eta_\lambda = \lambda \eta_\lambda,
\qquad
H_{r, \lambda} R_{E,\lambda} = E R_{E,\lambda}
\quad\hbox{with}\quad
H_{r,\lambda}= H_r + \lambda r^{-2}.
\label{1.7}\end{equation}
This is equivalent to specifying self-adjoint domains for the Hamiltonians
$H_\Omega$ and $H_{r,\lambda}$.

It is well-known (see also (\ref{A.9})) that for angular eigenvalues
$\lambda < \frac{\hbar^2}{8m} [3-(N-2)(N-4)]$
 the self-adjoint domain  of $H_{r,\lambda}$ is not
unique \cite{DS,Richt,Meetz}.
(For $N=3$ with the conventions (\ref{2.4}) adopted later
this means  $\lambda <1$.)
The spectra of the possible
self-adjoint versions of $H_{r,\lambda}$ have  been analyzed in a
recent paper \cite{indPLA}, which uses  the eigenvalues of
$H_\Omega$ provided by Calogero as input. What we do here is
different, since we aim to tackle the problem of constructing
inequivalent quantizations for $H_\Omega$ as well. On the one hand,
the inequivalent definitions  of the radial Hamiltonian correspond
to non-trivial contact interactions when {\em all}  particles
collide with each other (at the $r=0$ location). On the other hand,
the inequivalent self-adjoint domains of $H_\Omega$ can be obtained
by imposing different admissible boundary conditions for the angular
wave function at the singularities of the potential term in
(\ref{1.6}), which occur at the collisions of {\em any pairs} of
particles. This latter poses a more interesting problem to us,  but
it is also more difficult technically. We shall stick to the $N=3$
case for which the angular Schr\"odinger equation is still an
ordinary differential equation on the circle $S^1$.

The content of the present
paper and our main results can be outlined as follows.
After fixing the general framework and conventions in Section 2,
the self-adjoint extensions of the angular Hamiltonian are
described in Section 3
(with some details deferred to Appendix A).
It is  explained that the most general
local, self-adjoint boundary condition is given by eq.~(\ref{3.14}),
where $U_{\x}\in U(2)$
determines the connection condition for the
wave function at each of the six coincidence angles,
 ${\x}\in\cS$ (\ref{3.5}),  of two particle
positions (see Figure 1).
We then show that the dihedral $D_6$ symmetry of the angular potential,
which includes the $S_3$ subgroup given by the
permutations of the 3 particles, is maintained if
the connection condition is chosen uniformly with $U_{\x}=U$
independent of ${\x}$, subject to (\ref{3.15}).
The so-arising 2-parameter family of angular boundary conditions,
defined by $U$ in (\ref{3.18}),
has not been considered before for the $N=3$ Calogero model.
Calogero's study \cite{Cal3} corresponds, in effect,
to the $U=-{\bf 1}_2$ case.

Section 4 is the heart of the paper, containing a detailed
description of the eigenstates of the angular
Hamiltonian based on their classification under the $D_6$ symmetry.
There are two qualitatively different cases, according to whether
$U$ in (\ref{3.18}) is diagonal or non-diagonal.
In the former case the eigenfunctions over the angular
sectors corresponding to the various orderings of the particles
can be chosen independently, like in Calogero's original quantization.
If $U$ is non-diagonal, then the sectors are connected since
the probability current associated with the angular Hamiltonian
does not vanish in general at the coinciding particle positions.
It turns out that the angular eigenfunctions can be written
down explicitly once the eigenvalues are determined.
The eigenvalues are shown to have the form $\lambda = (3\mu)^2$,
where $\mu$ is any real or purely
imaginary solution of eqs.~(\ref{5.2})-(\ref{5.4}).
These 6 equations (counting the signs) are associated  with
the 6 inequivalent irreducible representations of $D_6$, which are
summarized in Appendix B.
Note that $D_6$ admits 1-dimensional
and  2-dimensional irreducible representations and the
associated eigenstates
and eigenvalues are called type 1 and type 2, respectively.

Because of the $D_6$ symmetry of the Hamiltonian,
it is consistent to regard the 3 interacting particles
as indistinguishable.
The 6 inequivalent representations of $D_6$ become pairwise
equivalent upon reduction to the $S_3 \subset D_3$ subgroup
generating the permutations of the 3 particles.
Thus the Hilbert space of the system
decomposes as an orthogonal direct sum of 3
subspaces containing, respectively, the exchange even (bosonic) states,
the exchange odd (fermionic) states,  and the states
associated with the $2$-dimensional representation of $S_3$
(that corresponds to parastatistics).
On the basis of cluster separability \cite{Peres}
or the assumption of a complete set of commuting observables
\cite{GP}, one may truncate the Hilbert space
to its bosonic or fermionic subspace in physical applications.
We stress that our inequivalent quantizations are new even after
such truncations.
The question of statistics in Calogero models is an  intriguing issue,
as is clear from \cite{Poly1}  relating  the model
(for $\omega=0$ in (\ref{1.1})) to fractional statistics.
See also the review \cite{Poly2} and references therein.

The angular eigenvalue equations (\ref{5.2})-(\ref{5.4})
cannot be solved explicitly for general $U$, and
Section 5 is devoted to the qualitative analysis of their solutions.
We provide (with some details contained in Appendix C) a characterization
of the shape of the functions entering these
equations,  as illustrated by Figures 2 and 3, which
allows us to see the general features of the angular
eigenvalues.
It turns out that, in addition to the always existing
unbounded infinite series of positive
eigenvalues, for certain boundary conditions
the angular Hamiltonian possesses
a finite number of negative eigenvalues, too.
In Section 5.4 some remarks are given concerning the
stability of the boundary conditions
admitting a negative eigenvalue and similarly for
the ones  possessing purely positive spectra.

The angular boundary conditions admitting a negative eigenvalue
are physically not
permissible, since if $\lambda<0$ then the spectrum of the
radial  Hamiltonian (\ref{6.2}) is not bounded from below.
This is demonstrated in Section 6, where we also describe the 1-parameter
family of the radial boundary conditions that arise for $\lambda <1$,
and characterize the corresponding
energy spectra qualitatively.
The eigenvalues of the radial Hamiltonian are found as the solutions
of equation (\ref{6.16}), where the shape of the function $F_\lambda$
is proved to be as illustrated by Figure 4.
The results derived in this section are consistent with the
previous analysis of the radial Hamiltonian \cite{indPLA},
which used a different method and adopted the
(then always positive)
angular eigenvalues from Calogero's paper.

We can explicitly write down all eigenvalues and eigenvectors
of the angular Hamiltonian in the four special cases
$U=\pm {\bf 1}_2, \pm \sigma_1$.
The complete solution of the $N=3$
Calogero model in these cases is displayed in Section 7.
As already mentioned Calogero's solution corresponds to
$U=-{\bf 1}_2$, but we argue that the $U=\sigma_1$ case is in a
sense more natural, since under this boundary condition
all eigenvalues and eigenstates
are smoothly connected to those of
the standard 2-dimensional harmonic oscillator
arising from $H_{rel}$ (\ref{1.4})-(\ref{1.6})  in the $g\to 0$ limit.

Some further comments on our results and
on open problems are given in Section 8.

\section{Separation of variables}
 \setcounter{equation}{0}

Although we will consider in detail the $N=3$ case only,
is worth explaining our strategy to construct a self-adjoint operator
from the formal expression $H_{rel}$ (\ref{1.4}) in general terms.

The first step is to choose a domain for the formal operator
$H_\Omega$ so that it yields
a self-adjoint operator of the Hilbert
space $L^2(S^{N-2})$, which we denote here by ${\hat H}_\Omega$.
We assume that
$L^2(S^{N-2})$ can be decomposed into a direct sum of the
eigensubspaces of ${\hat H}_\Omega$,
\begin{equation}
L^2(S^{N-2})= \oplus_\lambda V_\lambda,
\label{2.1}\end{equation}
where $\lambda$ is the corresponding eigenvalue.
Since $L^2({\bf R}^{N-1})= L^2(\bR_+, r^{N-2} dr) \otimes L^2(S^{N-2})$,
the direct sum
 (\ref{2.1}) induces  the decomposition
\begin{equation}
L^2(\bR^{N-1})= \oplus_\lambda\,
L^2(\bR_+, r^{N-2}dr)  \otimes  V_\lambda.
\label{2.2}\end{equation}
The next step is to construct a
self-adjoint  radial Hamiltonian,  $ {\hat H}_{r,\lambda}$,  of the Hilbert
space  $L^2(\bR_+, r^{N-2}dr)$
 out of the formal expression $H_{r,\lambda}$ that appears in (\ref{1.7}).
 Finally, we obtain a
 self-adjoint version of the formal relative Hamiltonian $H_{rel}$
(\ref{1.4}) by the   infinite direct sum
   \begin{equation}
  {\hat H}_{rel} = \oplus_\lambda
{\hat H}_{r,\lambda}\otimes \mathrm{id}_{V_\lambda},
\label{2.3}\end{equation}
in correspondence with the direct sum decomposition (\ref{2.2}) of the
Hilbert space of our system.

Let us now specialize to $N=3$.
Following  closely the lines of Calogero \cite{Cal3} we set
\begin{equation}
\hbar=2m =  1,
\label{2.4}\end{equation}
and introduce polar coordinates $(r,\phi)$ on the reduced configuration
space  $\bR^2$
 in such a way that we have
\begin{eqnarray}
&& x_1 - x_2 = r \sqrt{2} \sin \phi \nonumber\\
&& x_2 - x_3 =  r \sqrt{2} \sin( \phi + \frac{2}{3}\pi) \nonumber\\
&& x_3 - x_1 =  r \sqrt{2} \sin( \phi + \frac{4}{3}\pi).
\label{2.5}\end{eqnarray}
The angle  $\phi$ counts modulo $2\pi$ and $r\geq 0$.
The angular Hamiltonian acts on functions on $S^1$ by
\begin{equation}
M:= H_\Omega = - \frac{ d^2} { d \phi^2}  + \frac{g}{2} \frac{9}{\sin^2 3\phi}.
\label{2.6}\end{equation}
The radial Hamiltonian associated with an eigenvalue $\lambda$ of $M$
acts on functions on $\bR_+$ by
\begin{equation}
H_{r,\lambda}= -   \frac{d^2 }{d r^2}  - \frac{1}{r} \frac{d}{dr} +
 \frac{3}{8} \omega^2 r^2 + \frac{\lambda}{r^2}.
\label{2.7}\end{equation}
Because of (\ref{2.4}), our assumption (\ref{1.3})
on the range of $g$ becomes
\begin{equation}
-\frac{1}{2} < g < \frac{3}{2},
\label{2.8}\end{equation}
and it will be convenient to parametrize  $g$ as
\begin{equation}
g = 2\nu (\nu -1)
\quad\hbox{with}\quad
\frac{1}{2}< \nu < \frac{3}{2},
\quad (\nu \neq 1).
\label{2.9}\end{equation}

In the above we have
assumed that the eigenvectors of ${\hat H}_\Omega$ form a
complete set yielding the decomposition (\ref{2.1}).
We shall see later that this
assumption is satisfied for all self-adjoint angular Hamiltonians if $N=3$,
since these operators all have pure discrete spectra.
As for the radial Hamiltonian ${\hat H}_{r,\lambda}$, the
discreteness of its spectrum follows from general theorems for any $N$.
This implies, by (\ref{2.2}),  that
the eigenvectors of  ${\hat H}_{rel}$ (\ref{2.3})
obtained by the separation of variables span the Hilbert space
$L^2({\bf R}^{N-1})$ whenever ${\hat H}_\Omega$ admits a complete
set of eigenvectors in $L^2(S^{N-2})$.
In the subsequent sections we describe
the possible operators ${\hat H}_\Omega$ and ${\hat H}_{r,\lambda}$
that arise for $N=3$, and characterize
their spectra and  their eigenvectors.

\section{The definition of the angular Hamiltonian $M^U$}
\setcounter{equation}{0}

We now begin to study the angular Hamiltonian $M$  given by
(\ref{2.6}) with (\ref{2.9}).
At the formal level, the differential operator $M$ admits $D_6$
symmetry, i.e.,
the geometric symmetry of the regular hexagon.
We wish to maintain this symmetry in our quantization of the Calogero model,
and for this reason we select a self-adjoint domain for $M$ which
is left invariant under the $D_6$ transformations.
The physical motivation for the $D_6$ symmetry  comes from
the following two assumptions.
First, the $3$ particles be identical.
Second, the pairwise collisions of $2$ particles be equivalent
for the situations when the
spectator particle that does not participate in the collision
is located to the left or to the right of the point of collision
on the line.
Technically,
the first assumption leads to the usual $S_3$ symmetry group generated
by the exchanges of the particles, while the addition of the second
assumption renders the total symmetry group to be $D_6$ containing
the $S_3$ as a subgroup.
In other words,
the symmetry generators of $D_6$
that are not in the `exchange-$S_3$' subgroup are required
in order to ensure  the identical nature of those  singular points of $M$
at the quantum level which are not related by the particle exchanges,
like the singular points at $\phi=0$ and at $\phi=\pi$ (see Figure 1).

\begin{figure}[ht]  \centering  
\resizebox{.6\textwidth}{!}{\includegraphics{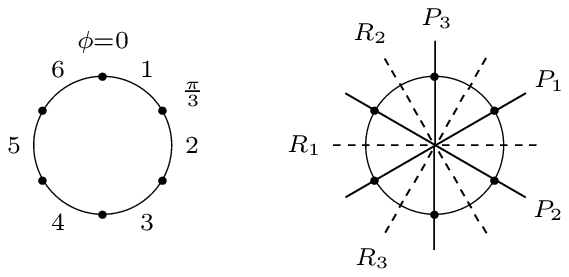}}
\caption{
The angular configuration space $S^1$, with the six singular points
located at $\cS$ (\ref{3.5}) and the six `sectors' between the
consecutive singularities (left), and with the
axes of the reflection symmetries of the angular Hamiltonian (right).
}
\label{fig:circles}  \end{figure}

\medskip

Referring  to Figure \ref{fig:circles},  let us note that the dihedral group
$D_6$ is generated by  the reflections $P_3$ and $R_3$
that operate on the circle as
\begin{equation}
P_3: \phi \mapsto - \phi,
\qquad
R_3:  \phi \mapsto \frac{\pi}{3} - \phi
\qquad (\mathrm{mod}\,\, 2\pi ).
\label{3.1}\end{equation}
Their composition is rotation by $\frac{\pi}{3}$,
\begin{equation}
\cR_{\frac{\pi}{3}}= R_3 \circ P_3: \phi \mapsto \phi + \frac{\pi}{3},
\label{3.2}\end{equation}
which generates the cyclic subgroup $C_6 \subset D_6$.
The other reflection elements are provided by
\begin{equation}
P_n = (\cR_{\frac{\pi}{3}})^n \circ P_3\circ  (\cR_{\frac{\pi}{3}})^{-n},
\qquad
R_n = (\cR_{\frac{ \pi}{3}})^n \circ R_3\circ (\cR_{\frac{\pi}{3}})^{-n}
\quad\hbox{for}\quad n=1,2.
\label{3.3}\end{equation}
The reflections $P_i$ generate an $S_3$ subgroup of $D_6$, which we call the
`exchange-$S_3$', since it acts by permuting the original particle positions.
For example, $P_3$ exchanges $x_1$ and $x_2$  (\ref{2.5}).
The reflections $R_i$  generate another $S_3$ subgroup,
which we call the `mirror-$S_3$' in what follows.
The intersection of these two $S_3$ subgroups is generated by the
cyclic permutation, $\cC$,
of the particles  given by
\begin{equation}
\cC := (\cR_{\frac{\pi}{3}})^{-2}.
\label{3.4}\end{equation}
The $D_6$ transformations map to itself the set of angles, $\cS$,
corresponding to the singular points of $M$
\begin{equation}
\cS= \{  \frac{k \pi}{3}\,\vert\, k=0,1,2,3,4,5\}.
\label{3.5}\end{equation}
For any function $\psi$ on $S^1$ or on $S^1\setminus \cS$
and $g\in D_6$, define the function $\hat g \psi$ by
\begin{equation}
(\hat g \psi)(\phi):= \psi( g^{-1}(\phi)).
\label{3.6}\end{equation}
Obviously, $\hat g$ is a unitary operator on the angular Hilbert
space $L^2(S^1)$.
It will yield  a symmetry  if it maps the chosen self-adjoint
domain of $M$ to itself.

We apply a fairly well-known procedure
(described in great detail for example in \cite{DS})
to specify self-adjoint domains for
the angular Hamiltonian.
As usual,  we denote the differential operator $M$  applied on some
domain $\cD$ by $M_\cD$,
and start by considering  the `minimal domain'  ${\cal D}_0$ consisting of
$C^\infty$
complex functions on $S^1\setminus \cS$ with compact support.
The domain of the adjoint $M_{\cD_0}^+$  of $M_{\cD_0}$ is,  in fact,
 the maximal domain $\cD_1 \subset L^2(S^1)$ on which $M$
can act as a differential operator.
This means that  $\cD_1$ consists of those complex functions  $\psi$
on $S^1\setminus \cS$  that together with their
first derivatives are absolutely continuous on any closed
interval (in $\phi$)
contained in $S^1\setminus \cS$
and for which both $\psi$ and $M\psi$ belong to $L^2(S^1)$.
It is a standard matter to show that $M_{\cD_0}^+= M_{\cD_1}$.
Therefore  $M_{\cD_0}$ is  a symmetric operator, and its self-adjoint
extensions
are restrictions of $M_{\cD_1}$ that can be obtained by imposing
suitable boundary conditions on the wave functions at the singular
points of the
 potential.

Note that the deficiency indices
\cite{DS,Richt,Weidmann} of $M_{\cD_0}$ are $(12,12)$
since any eigenfunction of $M$ on any of the six `sectors' on the circle
(see Figure \ref{fig:circles})
is square integrable due to (\ref{2.9}).
One can check this (well-known) result by means of the
explicit description of the eigenfunctions given in Section 4.1.
The facts that $M_{\cD_0}$ has finite deficiency indices and its self-adjoint
extension constructed by Calogero possesses pure discrete spectrum
permit one to conclude (see, e.g., Theorem 8.18 in \cite{Weidmann}) that
all self-adjoint extensions of $M_{\cD_0}$ possess
{\em pure discrete spectrum},
and thus also a complete set of eigenvectors in $L^2(S^1)$.

We shall describe  the self-adjoint boundary conditions in terms of certain
`reference modes' defined in pointed neighbourhoods of the elements of
$\cS$ (\ref{3.5}).
We first choose two reference modes around $\phi=0$, which we denote as
$\varphi^0_i$ for $i=1,2$. These are some {\em real} eigenfunctions of $M$
in some neighbourhood of $\phi=0$ (excluding $\phi=0$), normalized by
the Wronskian condition
\begin{equation}
W[\varphi_1^0,\varphi_2^0]:=
\varphi_1^0 \frac{d \varphi_2^0}{d\phi}-
\frac{d \varphi_1^0}{d\phi} \varphi_2^0 = 1.
\label{3.7}\end{equation}
After having chosen the reference modes around $0$,
we introduce reference modes $\varphi^{\x}_i$ around any ${\x}\in \cS$
by requiring that
\begin{equation}
\varphi_k^{R_i {\x}}(\phi) = (-1)^k \varphi_k^{\x}(R_i \phi)
\qquad \forall k=1,2,\quad i=1,2,3,\quad {\x}\in \cS.
\label{3.8}\end{equation}
This defines the $\varphi^{\x}_k$ uniquely.
We remark that
\begin{equation}
\varphi_k^{\cC {\x}}(\phi) = \varphi_k^{\x}( \cC^{-1} \phi).
\label{3.9}\end{equation}
Moreover, if the initial reference modes are chosen to satisfy
\begin{equation}
\varphi^0_k(-\phi) = (-1)^k \varphi_k^0(\phi),
\label{3.10}\end{equation}
then we also have
\begin{equation}
\varphi_k^{P_i {\x}}(\phi) = (-1)^k \varphi_k^\x(P_i \phi)
\quad\hbox{and}\quad
\varphi_k^{{\x}+\frac{\pi}{3}}(\phi) = \varphi_k^{\x}(\phi-\frac{\pi}{3}).
\label{3.11}\end{equation}

Using that the reference modes are  square integrable
due to (\ref{2.9}),  one can show
(for elementary arguments, see \cite{FTC,Krall})
that  the following `boundary vectors'  are well-defined for
any $\psi\in \cD_1$:
\begin{equation}
B_{\x}(\psi):= \left[\begin{array}{c}
    W[ \psi, \varphi_1^{\x}]_{{\x}+} \\
    W[ \psi, \varphi_1^{\x}]_{{\x}-}
\end{array}\right],
\quad
B'_{\x}(\psi):= \left[\begin{array}{c}
    W[ \psi, \varphi_2^{\x}]_{{\x}+} \\
   -W[ \psi, \varphi_2^{\x}]_{{\x}-}\end{array}\right]
   \quad\hbox{for}\quad {\x}=0, \frac{2\pi}{3}, \frac{4\pi}{3},
\label{3.12}\end{equation}
and
\begin{equation}
B_{\x}(\psi):= \left[\begin{array}{c}
    W[ \psi, \varphi_1^{\x}]_{{\x}-} \\
    W[ \psi, \varphi_1^{\x}]_{{\x}+}
\end{array}\right],
\quad
B'_{\x}(\psi):= \left[\begin{array}{c}
   - W[ \psi, \varphi_2^{\x}]_{{\x}-} \\
   W[ \psi, \varphi_2^{\x}]_{{\x}+}\end{array}\right]
   \quad\hbox{for}\quad {\x}= \frac{\pi}{3}, \pi, \frac{5\pi}{3}.
\label{3.13}\end{equation}
Here, $W[\psi, \varphi_i^{\x}]_{{\x}\pm}:=
\lim_{\epsilon \to \pm 0} W[\psi, \varphi_i^{\x}]({\x}+\epsilon)$.
According to the general theory of self-adjoint differential
operators \cite{DS},
the components of the boundary vectors give  a basis  of
the so-called  `boundary values'  for the
operator $M_{\cD_1}$,  and  the self-adjoint  boundary conditions
require the vanishing of
appropriate  linear combinations of the boundary values.

We restrict ourselves to  self-adjoint boundary conditions that
are local in the sense that they do not
mix the boundary values associated with  different singular points of $M$.
In fact
(see \cite{FTC,Kochubei,Gorba}  and Appendix A),
these boundary conditions
can be written as follows:
\begin{equation}
(U_{\x} - {\bf 1}_2) B_{\x}(\psi) + \ri  (U_{\x} + {\bf 1}_2) B'_{\x}(\psi)=0
\qquad \forall {\x}\in \cS,
\label{3.14}\end{equation}
where $U_{\x}\in U(2)$ are arbitrary unitary matrices.
This local boundary condition ensures that the quantum mechanical
probability current
on $S^1$ {\em remains continuous} at any point of $\cS$ for the
admissible wave functions
selected by  (\ref{3.14}).

It is important to observe that
{\em if $U_{\x}=U$ for all ${\x}\in \cS$ (\ref{3.5})
 and some $U\in U(2)$, then the boundary condition
(\ref{3.14}) is compatible with the mirror-$S_3$ symmetry.
If in addition the reference modes are chosen according to (\ref{3.10}) and
the `connection
matrix'  $U$ satisfies
\begin{equation}
U = \sigma_1 U \sigma_1,
\label{3.15}\end{equation}
then the boundary condition is compatible with
the full $D_6$ symmetry group.}

Indeed,
the first of the above statements is a consequence of the identities
\begin{equation}
B_{\x}( \hat R_i \psi) = B_{R_i {\x}}(\psi),
\qquad
B'_{\x}( \hat R_i \psi) = B'_{R_i {\x}}(\psi),
\label{3.16}\end{equation}
which are easily verified.
Under the assumption (\ref{3.10}), one also obtains the identities
\begin{equation}
B_{\x}( \hat P_i \psi) = \sigma_1 B_{P_i {\x}}(\psi),
\qquad
B'_{\x}( \hat P_i \psi) = \sigma_1 B'_{P_i {\x}}(\psi),
\label{3.17}\end{equation}
which imply the second statement.

The most  general $U\in U(2)$ subject to (\ref{3.15}) can be
written in the form
\begin{equation}
U= e^{\ri \alpha I} e^{\ri \beta \sigma_1} =
e^{\ri \alpha}
\left(  \begin{array}{cc}
  \cos \beta   &\ri \sin\beta   \\
  \ri \sin\beta  & \cos\beta
\end{array}\right)
:=\left(  \begin{array}{cc}
  \cA   &\cB   \\
  \cB & \cA
\end{array}\right).
\label{3.18}\end{equation}
The self-adjoint domains $\cD_U\subset \cD_1$ of our interest are
given by the boundary condition (\ref{3.14}) with $U_\theta=U$
in (\ref{3.18}) for all $\theta\in\cS$.
To denote  the  operator defined by
applying $M$ (\ref{2.6}) on the domain $\cD_U$, we use the notation
$M^U$ to exhibit the dependence on $U$.
In the next section we will fix the reference modes and  investigate
the  so-obtained
self-adjoint angular Hamiltonians in detail.
 The operator $M^U$ is the one denoted
by ${\hat H}_\Omega$ in Section 2, and henceforth the
`hat' is generally omitted from our self-adjoint operators for brevity.

In Section 7 the condition (\ref{3.14}) is referred to as
the `Dirichlet' case if  $U_{\x}=-{\bf 1}_2$ for all ${\x}\in \cS$,
the `Neumann' case
if $U_{\x}={\bf 1}_2$, and the `free' case if $U_{\x}=\sigma_1$.
To explain this terminology \cite{FTC}, note that the boundary condition
of the form (\ref{3.14}) can also be considered for sufficiently
regular potentials for which the
reference modes are smooth at ${\x}$ and can be chosen to satisfy
$\varphi_k^{\x}({\x}\pm 0)  = - \delta_{k,2}$ and
$\frac{d \varphi_k^{\x}} {d\phi} ({\x}\pm 0)  = \delta_{k,1}$.
In such circumstances (\ref{3.14}) becomes the standard
Dirichlet condition $\psi({\x})= 0$ if $U_{\x}=-{\bf 1}_2$, the Neumann condition
$\psi'({\x})=0$ if $U_{\x}={\bf 1}_2$, and the free boundary condition
requiring the continuity of $\psi$ and $\psi'$ at ${\x}$ if $U_{\x}=\sigma_1$.

\section{Eigenvalue equations and eigenvectors of $M^U$}
\setcounter{equation}{0}

In this section we develop a convenient formalism  to study the
eigenvalue-eigenvector equations for
 the angular Hamiltonian given by (\ref{2.6}), (\ref{2.9})
with the boundary condition (\ref{3.14}) specified by
$U_\theta=U$ from (\ref{3.18}) for all $\theta\in \cS$ (\ref{3.5}).
The cases of diagonal (`separating') and non-diagonal
(`non-separating') $U$ are considered separately.
In both cases the eigenvectors are obtained almost automatically
once the eigenvalues are determined, and they are classified
according to the representations of $D_6$ symmetry group (\ref{3.6}).
The eigenvalue equations  will be  further studied later.

\subsection{Preparations}

For any complex number $\mu$, consider the functions
\begin{eqnarray}
&& v_{1,\mu}(\phi) :=
\vert \sin 3\phi\vert^\nu F\left( \frac{\nu -\mu}{2}, \frac{\nu +\mu}{2}, \nu+
\frac{1}{2}; \sin^2 3\phi\right)
\nonumber \\
&& v_{2, \mu}(\phi) := \vert \sin 3\phi\vert^{1-\nu}
F\left( \frac{1-\nu -\mu}{2}, \frac{1-\nu +\mu}{2},- \nu+
\frac{3}{2}; \sin^2 3 \phi\right),
\label{4.1}\end{eqnarray}
where $F(a,b,c;z)$ is the standard hypergeometric function.
Locally,  the $v_{i,\mu}$ are linearly independent\footnote{The independence
requires $(\nu +\frac{1}{2})\notin \bZ$,  which explains why
 $g=-\frac{1}{2}$  is excluded from our treatment  (\ref{2.9}).}
 eigenfunctions of the
differential operator $M$  (\ref{2.6})
with eigenvalue $9 \mu^2$ \cite{Cal3}.
They are  square integrable
in  neighbourhoods of the singularities (\ref{3.5}),
since $\nu$ satisfies (\ref{2.9}).
  Unfortunately, these functions are not differentiable at those
values of $\phi$ for which
$\sin^2 3\phi =1$, where there is no singularity of the potential
in (\ref{2.6}),
and also do not satisfy the boundary condition in general.
Essentially, our problem is to select those linear combinations of
the $v_{i,\mu}$ that are smooth
on $S^1\setminus \cS$ and satisfy (\ref{3.14}).
Since $M$ with (\ref{3.14}) is self-adjoint, we can restrict our
attention to real or purely imaginary
values of $\mu$, for which $9 \mu^2$ is real.
 Since $\pm \mu$  define the same eigenfunctions, we can take
$\mu$ to be either
 non-negative or of the form $\mathrm{i} \vert \mu\vert$.

Below, we shall need the limiting values
\begin{equation}
a_i(\mu):= \lim_{\phi \to \frac{\pi}{6} -0} v_{i,\mu}(\phi),
\qquad
b_i(\mu):= \lim_{\phi \to \frac{\pi}{6} -0}\pa_\phi  v_{i,\mu}(\phi).
\label{4.2}\end{equation}
Explicitly, we have
   \be  
   \setlength{\arraycolsep}{.4ex}  
   \begin{array}{rclcrcl}
   a_1(\mu) & = & {\displaystyle
   \frac{\Gamma(\nu +\frac{1}{2}) \Gamma(\frac{1}{2})}
   {\Gamma(\frac{\nu+1+\mu}{2}) \Gamma(\frac{\nu+1-\mu}{2})}, } &
   \hskip 4ex  
   & a_2(\mu) & = &
   {\displaystyle \frac{\Gamma(-\nu +\frac{3}{2}) \Gamma(\frac{1}{2})}
   {\Gamma(\frac{-\nu+2+\mu}{2}) \Gamma(\frac{-\nu+2-\mu}{2})}, } \\
   b_1(\mu) & = &
   {\displaystyle \frac{6\, \Gamma(\nu +\frac{1}{2}) \Gamma(\frac{1}{2})}
   {\Gamma(\frac{\nu+\mu}{2}) \Gamma(\frac{\nu-\mu}{2})}, } &
   \rule{0ex}{6.0ex}  
   & b_2(\mu) & = &
   {\displaystyle \frac{6\, \Gamma(-\nu +\frac{3}{2})
\Gamma(\frac{1}{2})}
   {\Gamma(\frac{-\nu+1+\mu}{2}) \Gamma(\frac{-\nu+1-\mu}{2}) }. }
   \end{array}
   \label{4.3}\ee

Now we fix the boundary condition by choosing the reference
modes to be
\begin{eqnarray}
&& \varphi_1^0(\phi)= ( 3 (2\nu -1))^{-\frac{1}{2}} v_{1,\mu_0}( \phi)
[\Theta(\phi) - \Theta(-\phi)],
\nonumber\\
&& \varphi_2^0(\phi) = -  3^{-\frac{1}{2}}
(2\nu -1)^{\frac{1}{2}} v_{2,\mu_0}( \phi),
\label{4.4}\end{eqnarray}
where $\phi$ may vary as $ -\frac{\pi}{6}<\phi\neq 0  < \frac{\pi}{6}$,
$\Theta$ is the usual step function,
and $\mu_0$ is an arbitrary real number.
(The value of $\mu_0$ does not affect the boundary condition;
 only the asymptotic behaviour of the reference modes around $0$ matters.)

{}From this point on we study the operator $M^U:= M_{\cD_U}$, where
the self-adjoint domain
$\cD_U$ is specified by  (\ref{3.14})
using a matrix $U$ from (\ref{3.18}) and the above reference modes.
Note that eq.~(\ref{3.10}) holds, and thus $M^U$
admits the $D_6$ symmetry.

It is clear that the eigenfunctions of
$M^U$ yield smooth functions on $S^1\setminus \cS$.
In order to find  them, we adopt the following
strategy.
We first write down all eigenfunctions of the differential operator
$M$ that are smooth on
$S^1 \setminus \cS$.
We then select the admissible eigenvalues and eigenfunctions
by imposing the `connection conditions' (\ref{3.14}).

 Let us define the functions $\eta^1_{\pm,\mu}$ on $S^1\setminus \cS$ by
\begin{equation}
\eta^1_{+,\mu}( \phi) =
\left\{
\begin{array}{ll}
 b_2(\mu) v_{1,\mu}(\phi) - b_1(\mu) v_{2,\mu}(\phi) &
\mbox{if $0<\phi \leq \frac{\pi}{6}$ mod $2\pi$}   \\
  b_2(\mu) v_{1,\mu}(\frac{\pi}{3}-\phi) -
 b_1(\mu) v_{2,\mu}(\frac{\pi}{3}-\phi) &
\mbox{if $ \frac{\pi}{6}\leq\phi <\frac{\pi}{3}$ mod $2\pi$}   \\
 0 & \mbox{otherwise}
\end{array}
\right.
\label{4.5}\end{equation}
and
\begin{equation}
\eta^1_{-,\mu}( \phi) =
\left\{
\begin{array}{ll}
 a_2(\mu) v_{1,\mu}(\phi) - a_1(\mu) v_{2,\mu}(\phi) &
\mbox{if $0<\phi \leq \frac{\pi}{6}$ mod $2\pi$}   \\
  -a_2(\mu) v_{1,\mu}(\frac{\pi}{3}-\phi) + a_1(\mu)
v_{2,\mu}(\frac{\pi}{3}-\phi) &
\mbox{if $ \frac{\pi}{6}\leq\phi <\frac{\pi}{3}$ mod $2\pi$}   \\
 0 & \mbox{otherwise}
\end{array}
\right.
\label{4.6}\end{equation}
The functions $\eta_{\pm,\mu}^1$ are supported on `sector 1'
on the circle (see Figure \ref{fig:circles}) ,
and enjoy
the symmetry property $\hat R_3 \eta_{\pm,\mu}^1 = \pm \eta_{\pm,\mu}^1$.
In correspondence with the other five sectors on $S^1$, we introduce the
rotated functions
$\eta_{\pm,\mu}^k$ by
\begin{equation}
\eta_{\pm,\mu}^k(\phi) = \eta_{\pm,\mu}^1(\phi - (k-1) \frac{\pi}{3}),
\quad\hbox{for}\quad
k=2,\ldots, 6.
\label{4.7}\end{equation}
All these functions belong to $C^\infty(S^1\setminus \cS)$ and are
eigenfunctions of $M$,
\begin{equation}
M \eta_{\pm,\mu}^k   = 9 \mu^2 \eta_{\pm,\mu}^k.
\label{4.8}\end{equation}
They are square integrable for any $\mu$.
The most general smooth
eigenfunction of $M$ on $S^1\setminus \cS$ can be written as
the linear combination
\begin{equation}
\eta_\mu(\phi) = \sum_{k=1}^6 \left( C_+^k \eta^k_{+,\mu}(\phi) +
C_-^k \eta^k_{-,\mu}(\phi)\right),
\label{4.9}\end{equation}
with arbitrary complex numbers $C^k_\pm$.

Our problem is to select those linear combinations (\ref{4.9}) that
are compatible with the boundary condition (\ref{3.14}).
To do this, we need to compute the boundary vectors $B_{\x}(\eta_\mu)$ and
$B'_{\x}(\eta_\mu)$  for ${\x}\in \cS$ (\ref{3.5}).
By using the standard formulae for the asymptotic behaviour  of the
hypergeometric function,
we easily find that
\begin{eqnarray}
&&B_0(\eta_\mu)=   (3(2\nu -1))^{\frac{1}{2}}
\left[\begin{array}{c}
 -C_+^1b_1(\mu) - C_-^1 a_1(\mu)   \\
   -C_+^6 b_1(\mu) + C_-^6 a_1(\mu)
\end{array}\right], \nonumber\\
&&B'_0(\eta_\mu)=  (3(2\nu -1))^{\frac{1}{2}}
\left[\begin{array}{c}
 C_+^1b_2(\mu) +C_-^1 a_2(\mu)   \\
   C_+^6 b_2(\mu) -C_-^6 a_2(\mu)
\end{array}\right].
\label{4.10}\end{eqnarray}
The boundary vectors at any ${\x}\in \cS$ are then obtained from the identities
\begin{equation}
B_{\cR_{\frac{\pi}{3}} {\x}}(\eta_\mu) = \sigma_1
B_{\x}(\eta_\mu \circ \cR_{\frac{\pi}{3}} ),
\qquad
B'_{\cR_{\frac{\pi}{3}} {\x}}(\eta_\mu) =
\sigma_1 B'_{\x}(\eta_\mu \circ \cR_{\frac{\pi}{3}} ),
\label{4.11}\end{equation}
which follow from (\ref{3.16}) and (\ref{3.17}).
For example,
for $\eta_\mu$ in (\ref{4.9})
\begin{equation}
\eta_\mu \circ \cR_{\frac{\pi}{3}} =
\sum_{k=1}^6 \left( C_+^{\hat k} \eta^k_{+,\mu} +C^{\hat k} _-\eta^k_{-,\mu}
\right)
\quad\hbox{with}\quad
\mbox{ $\hat k = k+1$ mod  6},
\label{4.12}\end{equation}
and thus
\begin{eqnarray}
&&B_{\frac{\pi}{3}}(\eta_\mu)=   (3(2\nu -1))^{\frac{1}{2}}
\left[\begin{array}{c}
 -C_+^1b_1(\mu) +C_-^1 a_1(\mu)   \\
   -C_+^2 b_1(\mu) -C_-^2 a_1(\mu)
\end{array}\right], \nonumber\\
&&B'_{\frac{\pi}{3}}(\eta_\mu)=    (3(2\nu -1))^{\frac{1}{2}}
\left[\begin{array}{c}
 C_+^1b_2(\mu) -C_-^1 a_2(\mu)   \\
   C_+^2 b_2(\mu) +C_-^2 a_2(\mu)
\end{array}\right].
\label{4.13}\end{eqnarray}
All other boundary vectors can be calculated similarly.

\subsection{The separating cases}

For reasons that will become clear shortly, we call `separating'
the cases for which $U$ is diagonal:
\begin{equation}
U=
\left(  \begin{array}{cc}
  e^{\ri \alpha}   & 0   \\
  0 & e^{\ri \alpha}
\end{array}\right).
\label{4.14}\end{equation}
To start,  notice that the constants $C^1_\pm$ appear only in the
upper component
of the boundary condition (\ref{3.14})  for ${\x}=0$ and in the
lower component for ${\x}=\frac{\pi}{3}$.
By adding and subtracting these two equations we obtain
 \begin{eqnarray}
 &&[\ri (e^{\ri \alpha} +1) a_2(\mu) - (e^{\ri \alpha}-1) a_1(\mu)] C_-^1=0 ,
 \nonumber\\
&&[\ri (e^{\ri \alpha} +1) b_2(\mu) - (e^{\ri \alpha}-1) b_1(\mu)] C_+^1=0.
\label{4.15}\end{eqnarray}
The equations for $C_\pm^k$ decouple for different values of $k$,
and they are the same for any $k$.

Let us now look for a special solution for which $C_\pm^k=0$ for $k\neq 1$ and
\begin{equation}
\vert C_+^1\vert^2 + \vert C_-^1\vert^2\neq 0.
\label{4.16}\end{equation}
Assume that
\begin{equation}
e^{2\ri \alpha}\neq 1.
\label{4.17}\end{equation}
If both $C_+^1$ and $C_-^1$ were non-vanishing, then we could conclude from
(\ref{4.15}) that
\begin{equation}
\frac{a_1(\mu)}{a_2(\mu)} =  \frac{b_1(\mu)}{b_2(\mu)},
\label{4.18}\end{equation}
but it is possible to check (see equation (\ref{extra}) below)
that this can never happen.
Hence we have the following two sets of solutions:
\begin{equation}
\hbox{case A:}\quad \frac{a_1(\mu)}{a_2(\mu)}=\cot \frac{\alpha}{2},
\qquad
\eta_\mu= \eta_{-,\mu}^1,
\label{4.19}\end{equation}
\begin{equation}
\hbox{case B:}\quad \frac{b_1(\mu)}{b_2(\mu)}=\cot \frac{\alpha}{2},
\qquad
\eta_\mu= \eta_{+,\mu}^1.
\label{4.20}\end{equation}
The nature of the solutions of these `eigenvalue equations'
will be analyzed later.
It is clear that the general solution corresponding to an
eigenvalue determined by
(\ref{4.19}) and by (\ref{4.20}) is given respectively by
\begin{equation}
\eta_\mu^{A }= \sum_{k=1}^6 C_-^k \eta_{-,\mu}^k
\qquad\hbox{and}\qquad
\eta_\mu^B = \sum_{k=1}^6 C_+^k \eta_{+,\mu}^k ,
\label{4.21}\end{equation}
with arbitrary coefficients $C_\pm^k$.

If  $e^{\ri \alpha}=-1$,
then we obtain the following two sets of solutions:
\begin{equation}
\hbox{case A:}\quad  a_1(\mu)=0
\qquad\hbox{and}
\qquad
\hbox{case B:}\quad b_1(\mu)=0
\label{4.22-23}\end{equation}
with the associated eiqenfunctions
$\eta_\mu^A$ and $\eta_\mu^B$  of the same form as in (\ref{4.21}),
respectively.
In these cases the relevant formulae of $\eta^k_{\pm,\mu}$ simplify,
since in (\ref{4.5}) and (\ref{4.6})
only the contributions of $v_{1,\mu}$ survive.
The eigenvalues are given  according to
\begin{equation}
a_1(\mu)=0 \quad \leftrightarrow\quad
\mu^2= (2n+1 +\nu)^2,\qquad n=0,1,2,\ldots
\label{4.24}\end{equation}
\begin{equation}
 b_1(\mu)=0\quad \leftrightarrow\quad
\mu^2= (2n +\nu)^2,\qquad n=0,1,2,\ldots
\label{4.25}\end{equation}
This coincides with Calogero's solution \cite{Cal3} of the angular
Schr\"odinger equation for $N=3$.

If  $e^{\ri \alpha}=1$, then the following two sets of solutions result:
\begin{equation}
\hbox{case A:}\quad  a_2(\mu)=0 \qquad\hbox{and}\qquad
\hbox{case B:}\quad b_2(\mu)=0
\label{4.26-27}\end{equation}
with the corresponding
eigenfunctions provided by (\ref{4.21}).
Now only the contributions of $v_{2,\mu}$ survive in
$\eta^k_{\pm,\mu}$, and
the eigenvalues are furnished by
\begin{equation}
a_2(\mu)=0 \quad \leftrightarrow\quad
\mu^2= (2n+1 +(1-\nu))^2,\qquad n=0,1,2,\ldots
\label{4.28}\end{equation}
\begin{equation}
 b_2(\mu)=0\quad \leftrightarrow\quad
\mu^2=  (2n+(1-\nu))^2,\qquad n=0,1,2,\ldots
\label{4.29}\end{equation}

The boundary conditions with diagonal $U$ are `separating' in
the sense that they admit
a basis of  eigenfunctions  such that each eigenfunction  is supported on
a single sector on the circle between two singular points of
$M$ (\ref{2.6}).
In the separating case the multiplicity of each eigenvalue  is $6$,
in correspondence with the 6 arbitrary coefficients in the eigenfunctions
$\eta^A_\mu$
and $\eta^B_\mu$ above.
Let us denote by $\chi_\mu^A$ and $\chi_\mu^B$ the characters
of the representations of $D_6$ defined by (\ref{3.6})
on the eigensubspaces of $M^U$,
spanned by $\{\eta_{-,\mu}^k\}$ and $\{\eta_{+,\mu}^k\}$, respectively.
It is straightforward to calculate that
\begin{equation}
\chi^B_\mu(R_k)=-\chi^A_\mu(R_k) = 2,
\quad
\chi^X_\mu(P_k)= \chi^X_\mu(\cR_{\frac{\pi}{3}}^k)=0
\quad\hbox{for}\quad  k=1,2,3,\quad X=A,B.
\end{equation}
In terms of the irreducible characters described in Appendix B,
one obtains
\begin{equation}
\chi_\mu^A = \chi^{-+} + \chi^{--} + \chi^{(2)} + \tilde \chi^{(2)},
\qquad
\chi_\mu^B = \chi^{++} + \chi^{+-} + \chi^{(2)} + \tilde \chi^{(2)}.
\end{equation}
This fixes the decomposition of the eigensubspaces in the separating
case into irreducible representations of $D_6$.
If desired, one could easily
implement the decomposition explicitly.
The states associated with the characters $\chi^{\varrho +}$ are `bosonic'
and those associated with $\chi^{\varrho-}$ are `fermionic' under the
exchange-$S_3$ subgroup of $D_6$, for $\varrho=\pm$.
The explicit form of these bosonic and fermionic states is the same
that appears in eqs.~(\ref{4.55})-(\ref{4.56.1})  below.

\subsection{The non-separating cases}

Let us assume that $U$ in (\ref{3.18}) is non-diagonal.
Then we can rewrite the boundary condition for $\eta_\mu$ (\ref{4.9})
in the form
\begin{equation}
\left[
\begin{array}{c}
  C_+^{\hat k}   \\
  C_-^{\hat k}
\end{array}\right]
=T(\mu)
\left[
\begin{array}{c}
  C_+^{k}   \\
  C_-^{ k}
\end{array}\right]
\quad\hbox{with}\quad
\mbox{ $\hat k = k+1$ mod  6}, \quad \forall k=1,\ldots, 6.
\label{4.31}\end{equation}
The `transport matrix'  $T(\mu)$ is independent of $k$ because of the
$D_6$ symmetry of the problem.

To find $T(\mu)$, one may consider (\ref{3.14}) for $\psi=\eta_\mu$
with ${\x}=0$ and rewrite this equation as
 \begin{equation}
N_+(\mu) \left[ \begin{array}{c}
  C_+^{1}   \\
  C_-^{ 1}
\end{array}\right]
=
N_-(\mu)
\left[\begin{array}{c}
  C_+^{6}   \\
  C_-^{ 6}
\end{array}\right],
\label{4.32}\end{equation}
where
\begin{equation}
N_+=
\left[
\begin{array}{cc}
 (b - {\bar b}\cA) & (a-{\bar a} \cA)   \\
 -{\bar b}\cB & -{\bar a}\cB
\end{array}
\right],
\qquad
N_-=
\left[
\begin{array}{cc}
{\bar  b}\cB & -{\bar a} \cB   \\
 (-b +{\bar b}\cA) & (a -{\bar a}\cA)
\end{array}
\right],
\label{4.33}\end{equation}
and we have introduced the complex quantities
\begin{equation}
a:= a_1 +\ri a_2,
\qquad
b:= b_1 +\ri b_2.
\label{4.34}\end{equation}
These, and hence $N_\pm$, depend on $\mu$, but we dropped this from
the notation.
The $a_k(\mu)$ and $b_k(\mu)$ in (\ref{4.3}) are real,
since $\mu$ is real or purely imaginary,
and ${\bar a}$ and ${\bar b}$ are the complex conjugates of $a$ and $b$.
With the help of the reflection formula of the $\Gamma$-function,
we find the useful relations
\begin{equation}
a_1(\mu) b_2(\mu) - b_1(\mu) a_2(\mu)  = 3-6 \nu,
\quad
a_1(\mu) b_2(\mu)+ b_1(\mu)  a_2(\mu)  =
(3-6\nu) \frac{\cos\pi\mu}{\cos\pi\nu}.
\label{extra}\end{equation}
The first relation  implies that
\begin{equation}
\label{4.35}\det N_+ = \det N_-=(a {\bar b} - b {\bar a})\cB
=- 2\ri (3-6\nu)\cB ,
\end{equation}
which is non-zero precisely if $U$ is non-diagonal.
Thus we have
\begin{equation}
T(\mu) = N_+^{-1}(\mu) N_-(\mu),
\qquad
\det T(\mu)=1.
\label{4.36}\end{equation}
Explicitly, we obtain
\begin{equation}
T(\mu) = \frac{1}{\det N_+}
\left[
\begin{array}{cc}
 x(\mu) & y(\mu)   \\
 z(\mu) & x(\mu)
\end{array}
\right],
\label{4.37}\end{equation}
with
\begin{eqnarray}
&& x= -{\bar a} {\bar b}\cB^2 + ({\bar a} \cA -a)({\bar b}\cA -b),
\nonumber\\
&& y = {\bar a}^2 \cB^2 - ({\bar a}\cA -a)^2,
\nonumber\\
&&
z = {\bar b}^2 \cB^2 - ({\bar b}\cA -b)^2.
\label{4.38}\end{eqnarray}

We see from (\ref{4.31}) that any eigenstate is completely determined
if the corresponding coefficients $C_\pm^1$ are given, and hence
the dimension of the eigensubspaces is now at most 2.
 The $C_\pm^1$ and $\mu$ are of course
not arbitrary, the crudest condition on them being
\begin{equation}
T^6(\mu)
\left[
\begin{array}{c}
  C_+^{1}   \\
  C_-^{ 1}
\end{array}\right]
=
\left[
\begin{array}{c}
  C_+^{1}   \\
  C_-^{1}
\end{array}\right].
\label{4.40}\end{equation}
This condition must hold since the eigenfunctions of $M$ are
(smooth) functions
on $S^1\setminus \cS$, i.e.,
they enjoy the periodicity $\eta_\mu(\phi+2\pi)=\eta_\mu(\phi)$.
We simplify the requirement (\ref{4.40}) by classifying the eigenstates
according to the irreducible representations of the group $D_6$,
which is possible because
any operator $\hat g$ (\ref{3.6}) for $g\in D_6$ commutes with  $M^U$.
As summarized in Appendix B, the group $D_6$ admits four 1-dimensional
and two 2-dimensional irreducible representations.
In what follows,
we call an eigenstate `type 1' if it belongs to one of the 1-dimensional
representations, and `type 2' if it belongs to one of the
2-dimensional irreducible representations.
The corresponding
eigenvalues will be called type 1 and type 2 as well.
To describe the states in terms of these representations, it is convenient
to diagonalize, together with $M^U$, the operator
$\hat \cT$ that represents the
rotation\footnote{Our subsequent treatment  was inspired by \cite{Veigy},
where the cyclic permutation operator $\hat \cC=\hat \cT^2$
rather than $\hat \cT$ was used
to solve the system for $\frac{1}{2}< \nu < 1$ with
$U=\sigma_1$, in effect.}
\begin{equation}
\cT := (\cR_{\frac{\pi}{3}})^{-1}.
\label{notat}\end{equation}
The action  of $\hat\cT$ on a general eigenfunction (\ref{4.9})
is given by (\ref{4.12}):
\begin{equation}
\hbox{if}\quad
\eta_\mu = \sum_{k, \pm} C_\pm^k \eta_{\pm, \mu}^k,
\quad\hbox{then}\quad
{\hat \cT} \eta_\mu = \sum_{k, \pm} C_\pm^{\hat k} \eta^k_{\pm, \mu}
\quad\hbox{with}\quad
\mbox{ $\hat  k = k+1$ mod  6}.
\label{4.41}\end{equation}
Since $\hat \cT^6=\mathrm{id}$, the possible eigenvalues of $\hat \cT$
are the sixth roots of unity,
$e^{\frac{k\pi\ri}{3}}$, $k = 0, 1, \ldots, 5$, which are
$\pm 1,\, \pm \jmath,\, \pm \bar \jmath\,$ in terms of the cubic root
$\jmath := e^{\frac{2\pi\ri}{3}}$.
By combining (\ref{4.41})  with (\ref{4.31}), we see that a joint
eigenstate $\eta_\mu$ of $M^U$ and
$\hat \cT$, for which
\begin{equation}
{\hat \cT} \eta_\mu = \tau\, \eta_\mu,
\qquad \tau \in \{\,\pm 1,\, \pm \jmath,\, \pm \bar \jmath \, \},
\label{4.42}\end{equation}
is equivalently given by its coefficients $C_\pm^1$ that satisfy
\begin{equation}
T(\mu)
\left[
\begin{array}{c}
  C_+^{1}   \\
  C_-^{ 1}
\end{array}\right]
= \tau
\left[
\begin{array}{c}
  C_+^{1}   \\
  C_-^{1}
\end{array}\right].
\label{4.43}\end{equation}
This implies (\ref{4.40}) but is more useful to characterize the
eigenstates, because one can tell the `type' of the eigenstate from
the eigenvalue $\tau$: it is type 1 if $\tau = \pm 1$ and type 2 if
$\tau \in \{\, \pm \jmath,\, \pm \bar \jmath \, \}$
(see Appendix B).

The existence of a non-zero eigenvector in (\ref{4.43}) is equivalent to
\begin{equation}
0 = (\det N_+)^2 \det( T(\mu) - \tau \mathbf{1}_2)=
\left[ x(\mu) - \tau \det N_+\right]^2 - y(\mu)z(\mu).
\label{4.44}\end{equation}
Combining this with $\det T(\mu)=1$, and using (\ref{4.35})
and (\ref{4.38}) we find
\begin{equation}
\frac{x(\mu)}{\det N_+}=
\frac{({\bar a} \cA -a)({\bar b}\cA -b) -{\bar a} {\bar b}\cB^2}
{(a {\bar b} - b {\bar a})\cB}=
\Re(\tau),
\label{4.45}\end{equation}
where $\Re(\tau)$ denotes the real part of $\tau$.
For  $\tau$ in (\ref{4.42})
\begin{equation}
\Re(\tau)
=\frac{1 +\tau^2 }{2\tau }
= \left\{
\begin{array}{ll}
\pm 1 &  \mbox{if $\tau = \pm 1$,}   \\
\pm {1\over 2} & \mbox{if $\tau = \mp \jmath, \, \mp \bar\jmath$.}
\end{array}
\right.
\label{4.45.1}\end{equation}
The admissible values of $\mu$ are determined by the spectral
condition (\ref{4.45}) once $\tau$ is specified.

Let us now focus on the type 1 eigenstates for which $\tau = \pm 1$.
In this case the spectral
condition (\ref{4.45}) can be factorized as
\begin{equation}
\Bigl({\bar a} \cA  - a -  \Re(\tau) {\bar a} \cB \Bigr)
\Bigl({\bar b} \cA  - b +  \Re(\tau) {\bar b} \cB \Bigr)
= 0.
\label{4.45.2}\end{equation}
This allows us to classify the solutions according to the possibilities
as to which of the two factors in (\ref{4.45.2}) vanishes and which of
the signs $\Re(\tau) = \pm 1$ is chosen.
These four possibilities are listed as
\begin{equation}
\mbox{cases $\mathrm{A}_\pm$:}\quad
a(\mu) = (\cA \pm \cB)\, {\bar a}(\mu) ,
\label{4.47}\end{equation}
\begin{equation}
\mbox{cases $\mathrm{B}_\pm$:}\quad
b(\mu) = (\cA \pm \cB)\, {\bar b}(\mu) .
\label{4.48}\end{equation}
Note that we have assigned $\Re(\tau) = \mp 1$ to
$\mathrm{A}_\pm$ and
$\Re(\tau) = \pm 1$ to $\mathrm{B}_\pm$, respectively.
The conditions (\ref{4.47}), (\ref{4.48})
can be spelled out in the more explicit form
\begin{equation}
\mbox{cases $\mathrm{A}_\pm$:}\quad
\frac{a_1(\mu)}{a_2(\mu)}= \cot \frac{\alpha \pm \beta}{2},
\label{4.49}\end{equation}
\begin{equation}
\mbox{cases $\mathrm{B}_\pm$:}\quad
\frac{b_1(\mu)}{b_2(\mu)}= \cot \frac{\alpha \pm \beta}{2},
\label{4.50}\end{equation}
which
should be compared with (\ref{4.19}) and (\ref{4.20}).  We shall analyse
the roots of the eigenvalue equations (\ref{4.49}), (\ref{4.50}) later, and
here we just mention the special case in which
$\cot \frac{\alpha \pm \beta}{2}$ becomes divergent.
This happens precisely if
$(\cA\pm \cB)=e^{\ri (\alpha\pm \beta)}=1$, whereby the above
equations formally give
$a_2(\mu)=0$ and $b_2(\mu)=0$, respectively.
These conditions result also directly from (\ref{4.47}), (\ref{4.48}),
because they simplify as
\begin{equation}
a(\mu)=\bar a(\mu)\quad\hbox{and} \quad b(\mu)=\bar b(\mu)\quad\hbox{ if}
\quad
 e^{\ri (\alpha \pm \beta)}=1,
\label{4.51}\end{equation}
which can be solved trivially.
Similarly, we have
\begin{equation}
a(\mu)=-\bar a(\mu)\quad\hbox{and} \quad b(\mu)=
-\bar b(\mu)\quad\hbox{ if}\quad
 e^{\ri (\alpha \pm \beta)}=-1,
\label{4.52}\end{equation}
and then the solution is obtained from $a_1(\mu)=0$ and $b_1(\mu)=0$
according to (\ref{4.24}) and (\ref{4.25}).

Since the cases
$\mathrm{A}_\pm$ involve only $a(\mu)$ and $\bar a(\mu)$ while
$\mathrm{B}_\pm$ involve only $b(\mu)$ and $\bar b(\mu)$,
each of these conditions occurs if the eigenstate $\eta_\mu$ (\ref{4.9})
consists exclusively of $\eta_{-, \mu}^k$ or $\eta_{+, \mu}^k$
(see (\ref{4.5}) and (\ref{4.6})).
By using this observation and the assignment of
$\Re(\tau) = \mp 1$ to $\mathrm{A}_\pm$ and
$\Re(\tau) = \pm 1$ to $\mathrm{B}_\pm$,  we immediately
find the corresponding eigenstates to be
\begin{equation}
\eta_\mu^{A_+}= \sum_{k=1}^6 (-1)^{k+1} \eta_{-,\mu}^k,
\qquad
\qquad
\eta_\mu^{A_-}=
\sum_{k=1}^6 \eta_{-,\mu}^k,
\label{4.55}\end{equation}
\begin{equation}
\eta_\mu^{B_+}=
\sum_{k=1}^6  \eta_{+,\mu}^k,
\qquad
\qquad
\eta_\mu^{B_-}=
\sum_{k=1}^6 (-1)^{k+1} \eta_{+,\mu}^k.
\label{4.56}\end{equation}
One can check that the reflections $R_k, P_k\in D_6$ act
on these states according to
\begin{equation}
\hat R_k \eta_\mu^{A_\pm} = -  \eta_\mu^{A_\pm},
\qquad
\hat R_k \eta_\mu^{B_\pm} =   \eta_\mu^{B_\pm},
\qquad
\hat P_k \eta^{X_\pm}_\mu = \pm \eta^{X_\pm}_\mu\quad\hbox{for}
\quad X=A,B,
\label{parity}\end{equation}
which fixes their association with the four 1-dimensional
representations of $D_6$.
If we label the type 1 eigenstates by the parities analogously
to the labeling of the type 1 characters of $D_6$ in Appendix B,
then we can write
\begin{equation}
\eta_\mu^{A_+} = \eta_\mu^{-+},\qquad
\eta_\mu^{A_-} = \eta_\mu^{--},\qquad
\eta_\mu^{B_+} = \eta_\mu^{++},\qquad
\eta_\mu^{B_-} = \eta_\mu^{+-}.
\label{4.56.1}\end{equation}
In particular, the states $\eta^{X_+}_\mu$ are `bosonic' and the
$\eta^{X_-}_\mu$ are `fermionic' under the exchange-$S_3$.

The eigenstates (\ref{4.55}), (\ref{4.56}) may also be confirmed
by examining the eigenvectors in (\ref{4.43}).  In fact,
the conditions (\ref{4.47}) and (\ref{4.48}) are equivalent to
$y(\mu) = 0$ and $z(\mu) = 0$, respectively, and hence the
`transport matrix' becomes triangular for the type
1 eigenvalues,
\begin{equation}
\mbox{cases $\mathrm{A}_\pm$:}\quad
T(\mu) =
\left[\begin{array}{cc}
 \mp 1 &  0  \\
   \frac{z(\mu)}{\det N_+}  & \mp 1
\end{array}\right],
\label{4.53}\end{equation}
\begin{equation}
\mbox{cases $\mathrm{B}_\pm$:}\quad
T(\mu) =
\left[
\begin{array}{cc}
 \pm 1 &   \frac{y(\mu)}{\det N_+}   \\
 0 & \pm 1
\end{array}
\right].
\label{4.54}\end{equation}
The obvious eigenvector of $T(\mu)$ gives rise to
the corresponding eigenstate in (\ref{4.55}), (\ref{4.56}).
It is also possible to show that  $z(\mu)$ and $y(\mu)$ can never
vanish simultaneously, and thus $T(\mu)$ above
is truly  triangular.
This implies that the multiplicity of
each type 1 eigenvalue is 1.

Next, we turn to the type 2 eigenstates that span the 2-dimensional
eigensubspaces of $M^U$.
The states belonging to the $D_6$ representation of character
$\chi^{(2)}$ (see Appendix B) may be denoted as $\eta^{(2)}_{\mu,\tau}$
with $\tau = -\jmath$, $-\bar\jmath$ and those belonging to
the representation of character
$\tilde \chi^{(2)}$ as $\tilde\eta^{(2)}_{\mu,\tau}$
with $\tau = \jmath$, $\bar\jmath$.
For each $\Re(\tau) = \pm {1\over 2}$, the admissible values of
$\mu$ are determined by the spectral condition (\ref{4.45}),
which  can be expanded straightforwardly as
\begin{equation}
\frac{ \sin\alpha}{\sin \beta}  \frac{a_1 b_2 + a_2 b_1}{a_1 b_2 - a_2 b_1}
+ \frac{\cos\beta - \cos \alpha}{\sin \beta}
\frac{a_1 b_1}{a_2 b_1 - a_1 b_2}
+ \frac{\cos\beta + \cos \alpha}{\sin \beta} \frac{a_2 b_2}{a_2 b_1 - a_1 b_2}
=
\Re(\tau).
\label{4.62}\end{equation}
Upon substituting (\ref{extra}) into (\ref{4.62}), we obtain
\begin{equation}
\frac{ \sin\alpha}{\sin \beta}  \frac{\cos \pi \mu}{\cos  \pi \nu}
+ \frac{\cos\beta - \cos \alpha}{(6\nu -3) \sin\beta}  a_1(\mu) b_1 (\mu)
+ \frac{\cos\beta +  \cos \alpha}{(6\nu -3) \sin\beta}   a_2(\mu) b_2 (\mu)
=
\Re(\tau).
\label{4.63}\end{equation}
The products $a_1 b_1$ and $a_2 b_2$ do not simplify to trigonometric
functions.
The best we can do is to rewrite them using the identity
\begin{equation}
\Gamma(z) \Gamma(z+\frac{1}{2}) = \Gamma(2z) \Gamma(\frac{1}{2}) 2^{1-2z}
\label{4.64}\end{equation}
as
\begin{equation}
a_1(\mu) b_1(\mu) =  \frac{6\,\Gamma^2(\nu+\frac{1}{2})  2^{2(\nu -1)}}
{\Gamma(\nu+\mu) \Gamma(\nu -\mu)},
\qquad
a_2(\mu) b_2(\mu) = \frac{6\,\Gamma^2(-\nu+\frac{3}{2})  2^{-2\nu }}
{\Gamma(1-\nu+\mu) \Gamma(1-\nu -\mu)}.
\label{4.65}\end{equation}

In contrast to the type 1 case,
now some further work is needed to find
the eigenvectors in (\ref{4.43})
for any admissible $\mu$ solving (\ref{4.63}).
However, since there must be two independent eigenvectors
(with eigenvalues $\tau$ and $\bar\tau$)
to form a 2-dimensional representation of $D_6$,
the condition (\ref{4.40})
must actually hold automatically for arbitrary $C_\pm^1$.
 This implies that we have the matrix identity
$T^6(\mu) = \mathbf{1}_2$ for any type 2
eigenvalue (as can also be confirmed by a direct computation).
Thus, we can associate with
any eigenvalue $\tau$ of $T(\mu)$ the projection operator
\begin{equation}
\pi_{\tau}(\mu) = \frac{1}{6} \sum_{k = 1}^6 \bar \tau^k T^k(\mu),
\label{4.65.1}\end{equation}
which
satisfies
$T(\mu)\pi_{\tau}(\mu)  = \tau \pi_{\tau}(\mu)$,
$\pi_{\tau}^2(\mu) = \pi_{\tau}(\mu)$
by virtue of  $\tau \bar\tau = 1$.
Then the eigenvector in (\ref{4.43}) is provided by
\begin{equation}
\left[\begin{array}{c}
C^1_+\\
C_-^1
\end{array}\right]_{\tau}=\pi_{\tau}(\mu)
\left[\begin{array}{c}
\gamma \\
\delta
\end{array}\right],
\label{4.65.3}\end{equation}
with arbitrary complex numbers $\gamma$, $\delta$ for which the
eigenvector is non-vanishing.
The projection operators may be evaluated explicitly as
\begin{equation}
\pi_{\tau}(\mu) =
\left[\begin{array}{cc}
{1\over 2} &  \frac{-2\ri y(\mu)}{3\det N_+} \Im(\tau)  \\
\frac{-2\ri z(\mu)}{3\det N_+} \Im(\tau)  & {1\over 2}
\end{array}\right],
\label{4.65.4}\end{equation}
where $\Im(\tau)\in \{\pm \frac{\sqrt{3}}{2}\}$
denotes the imaginary part of $\tau$.
In principle, the construction is completed by using (\ref{4.31}) and
(\ref{4.9}) to generate  the type 2 eigenfunctions of $M^U$.

The eigenvalues of $M^U$ cannot be
presented explicitly in general.
In the next section we investigate some
features of the eigenvalues for general $U$ (\ref{3.18}).
We know of four explicitly solvable cases, corresponding to
$U = \pm\mathbf{1}_2$, $\pm \sigma_1$.
These are discussed in Section 7.

\section{Characterization of the eigenvalues of $M^U$}
\setcounter{equation}{0}

We have seen that the eigenvalues of the angular Hamiltonian $M^U$
are given  by
\begin{equation}
\lambda = (3 \mu)^2,
\label{5.1}\end{equation}
where $\mu$ is a solution of one of the following  `eigenvalue equations'.
First,
\begin{equation}
F_A(\mu):=
\frac{\Gamma(\frac{1+\nu+\mu}{2}) \Gamma(\frac{1+\nu-\mu}{2}) }
{ \Gamma(\frac{2-\nu+\mu}{2}) \Gamma(\frac{2-\nu-\mu}{2}) }
=
\frac{\Gamma(\nu +\frac{1}{2})}{\Gamma(-\nu +\frac{3}{2})}
\tan \frac{\alpha \pm \beta}{2}
\label{5.2}\end{equation}
or
\begin{equation}
F_B(\mu):=
\frac{\Gamma(\frac{\nu+\mu}{2}) \Gamma(\frac{\nu-\mu}{2}) }
{ \Gamma(\frac{1-\nu+\mu}{2}) \Gamma(\frac{1-\nu-\mu}{2}) }
=\frac{\Gamma(\nu +\frac{1}{2})}{\Gamma(-\nu +\frac{3}{2})}
\tan \frac{\alpha \pm \beta}{2} .
\label{5.3}\end{equation}
Second,
\begin{equation}
F_2(\mu):= \frac{ \sin\alpha}{\sin \beta}  \frac{cos \pi \mu}{\cos  \pi \nu}
+ \frac{\cos\beta - \cos \alpha}{(6\nu -3) \sin\beta}  a_1(\mu) b_1 (\mu)
+  \frac{\cos\beta + \cos \alpha}{(6\nu -3) \sin\beta}   a_2(\mu) b_2 (\mu)
=\pm \frac{1}{2}.
\label{5.4}\end{equation}
Equations (\ref{5.2}) and (\ref{5.3}) control all eigenvalues
for the separating boundary conditions,
for which $\beta = 0$ modulo $\pi$, and the `type 1' eigenvalues
for the non-separating boundary
conditions. In the latter case,  the states corresponding to the
solutions of (\ref{5.2}) are odd
while those corresponding to (\ref{5.3}) are  even with respect to
the  mirror-$S_3$  symmetry.
Their parity under the exchange-$S_3$ symmetry is given  by the $\pm$
in the argument of
the tangent function on the right hand side.
Equation (\ref{5.4})  governs the `type 2' eigenvalues for
the non-separating boundary conditions.
In each of the cases, we are interested in the values of $\mu$ that
satisfy  either $\mu= \vert \mu\vert $ or
$\mu = \ri \vert \mu \vert$ in correspondence with the non-negative
and the negative eigenvalues of $M^U$.
In fact, one of the main questions is whether negative eigenvalues
exist or not.

In our subsequent analysis of equations (\ref{5.2}) and (\ref{5.3})
we assume that $\vert \tan \frac{\alpha \pm \beta}{2}\vert < \infty$.
If  $\tan \frac{\alpha \pm \beta}{2}$ diverges or vanishes, then
all solutions of the corresponding equations
are real and can be written down explicitly, as discussed
in Sections 4 and  7.

\subsection{Type 1  negative eigenvalues}

We now study the possibility of negative eigenvalues arising from
equations (\ref{5.2}), (\ref{5.3}), which govern
the type 1 eigenvalues  for non-separating $U$ {\em and}  all
eigenvalues for separating $U$.
By setting $\mu = \ri x$ with  $x\geq 0$, we  consider the real
functions on $ [0,\infty)$ defined  by
\begin{equation}
 x\mapsto F_A(\ri x)=
\frac{\vert \Gamma(\frac{1+\nu+\ri x }{2})\vert^2}
{ \vert \Gamma(\frac{2-\nu+\ri x }{2})\vert^2 }
\qquad\hbox{and}\qquad
 x\mapsto F_B(\ri x)=
\frac{\vert \Gamma(\frac{\nu+\ri x }{2})\vert^2}
{ \vert \Gamma(\frac{1-\nu+\ri x }{2})\vert^2 }.
\label{5.5}\end{equation}
We show below that these functions are strictly monotonically
{\em increasing}  on  $[0,\infty )$
and they grow {\em without any bound} as $x$ tends to $+\infty$.
 Therefore there exists {\em at most $1$} negative eigenvalue of $M^U$
that arises as a solution of
  (\ref{5.2}), for any fixed sign $\pm$  on the right hand side,
 and such eigenvalue occurs {\em precisely  if}  the parameters of
$M^U$ satisfy
   \begin{equation}
 F_A(0)= \frac{\vert \Gamma(\frac{1+\nu }{2})\vert^2}
{ \vert \Gamma(\frac{2-\nu}{2})\vert^2 } <
\frac{\Gamma(\nu +\frac{1}{2})}{\Gamma(-\nu +\frac{3}{2})}
\tan \frac{\alpha \pm \beta}{2}.
\label{5.6}\end{equation}
Similarly, for any fixed sign $\pm$ on the right hand side of (\ref{5.3}),
we obtain  at most 1 negative eigenvalue,  which occurs  if and only if
\begin{equation}
\ F_B(0)= \frac{\vert \Gamma(\frac{\nu }{2})\vert^2 }
{ \vert \Gamma(\frac{1-\nu}{2})\vert^2 } <
\frac{\Gamma(\nu +\frac{1}{2})}{\Gamma(-\nu +\frac{3}{2})}
\tan \frac{\alpha \pm \beta}{2} .
\label{5.7}\end{equation}

 To prove the above claims regarding  $F_A(\ri x)$, we inspect  its
logarithmic  derivative:
\begin{equation}
\frac{2}{\ri} \frac{d \log F_A(\ri  x) }{ d x}=
\psi\left(\frac{1+\nu + \ri x}{2}\right)
-\psi\left(\frac{1+\nu - \ri x}{2}\right)
+ \psi\left(\frac{2-\nu - \ri x}{2}\right)
-\psi\left(\frac{2-\nu + \ri x}{2}\right)
\label{5.8}\end{equation}
with the standard notation
\begin{equation}
\psi(z)= \frac{\Gamma'(z)}{\Gamma(z)}.
\label{5.9}\end{equation}
Recall the identity (\cite{GR}: 8.363 4.), for any real $\xi$ and $\eta$,
\begin{equation}
\psi(\xi+\ri \eta) - \psi(\xi - \ri \eta) =
\sum_{k=0}^\infty \frac{2\eta\ri}{\eta^2 + (\xi+k)^2}.
\label{5.10}\end{equation}
By using this and adding the two series, we obtain
\begin{equation}
\frac{d \log F_A(\ri x) }{ d x}= \sum_{k=0}^\infty
\frac{2x (2\nu -1) (3+4k)}
{[x^2 + (2-\nu +2k)^2] [ x^2 + (1+\nu + 2k)^2]},
\label{5.11} \end{equation}
which is positive for $x>0$ as  $(2\nu -1)>0$.
Since $F_A(\ri  x)>0$, this implies that $x\mapsto F_A(\ri x)$
is a strictly increasing function
on $[0, \infty)$. Next we recall (\cite{GR}: 8.328) that
\begin{equation}
\lim_{\vert \eta\vert \to \infty} \vert \Gamma(\xi+\ri \eta)
\vert e^{\frac{\pi}{2}\vert \eta\vert }
\vert \eta\vert^{\frac{1}{2}-\xi} = \sqrt{2\pi}.
\label{5.12}\end{equation}
It follows from (\ref{5.12}) and (\ref{2.9}) that
\begin{equation}
\lim_{x\to \infty} F_A(\ri x)= \lim_{x\to \infty}
\left(\frac{x}{2}\right)^{2\nu -1}= \infty,
\label{5.13}\end{equation}
as had been claimed.  The analogous properties of $F_B(\ri x)$ can be
verified in the same manner.

\subsection{Type 1 non-negative  eigenvalues}

\begin{figure}[ht]  \centering  
\resizebox{.35\columnwidth}{!}{\includegraphics{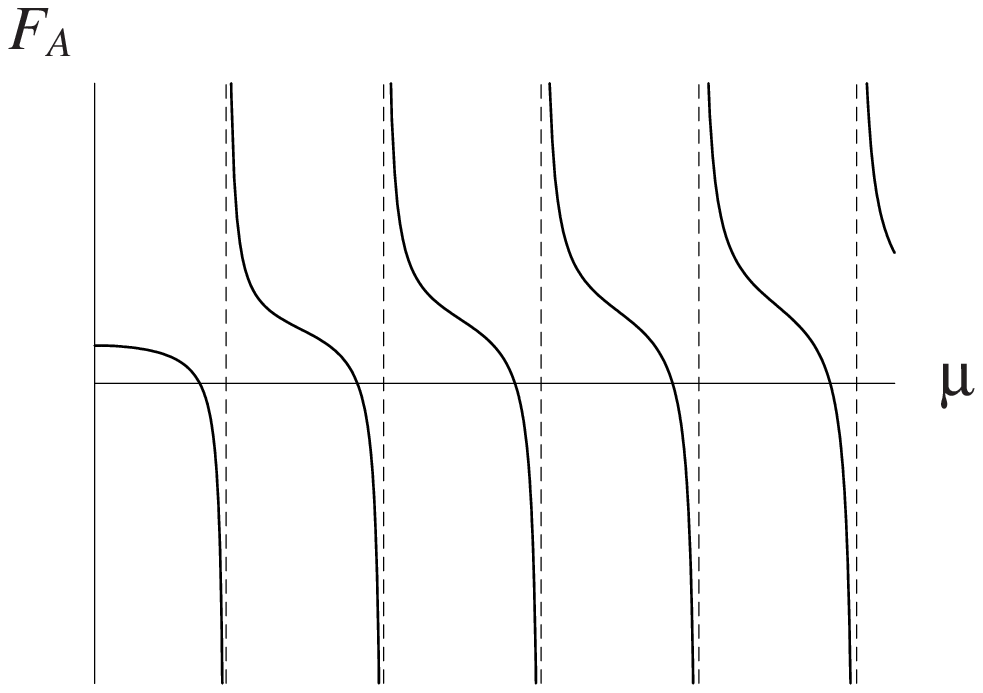}}
    \hskip .15\columnwidth
\resizebox{.35\columnwidth}{!}{\includegraphics{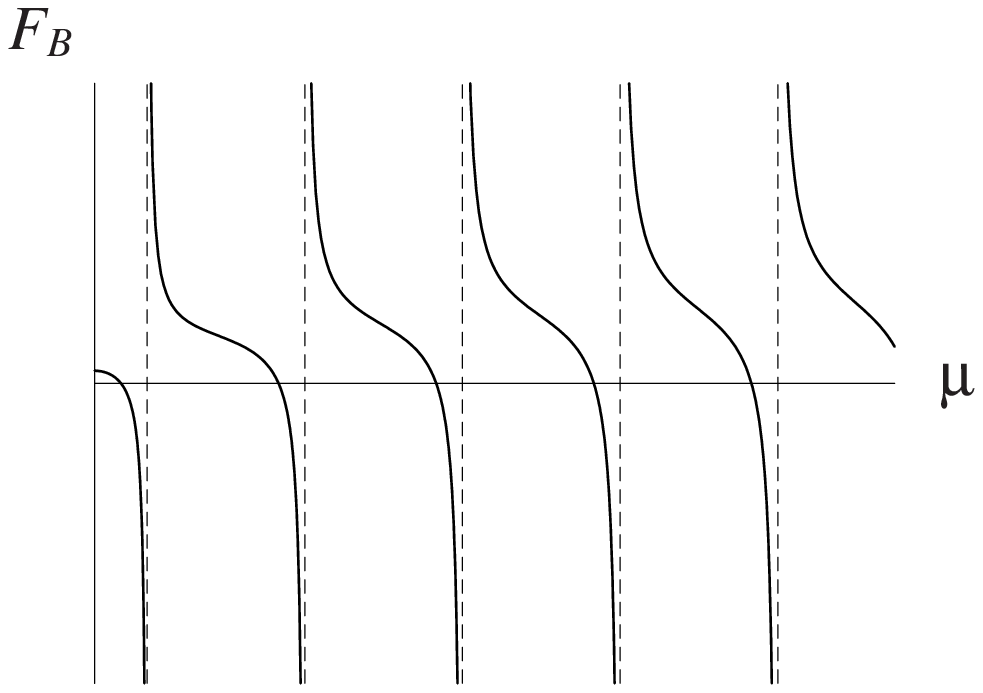}}
\caption{$F_A$ (left) and $F_B$ (right), as the function of $\mu$
[see (\ref{5.2})--(\ref{5.3})], for $\mu \ge 0$, with $\nu = 2/3$. The
dashed lines are located at $\mu_m^\infty$ (\ref{5.14}) for $F_A$ and
$\bar{\mu}_m^\infty$ (\ref{5.19}) for $F_B$.}
\label{fig:angular}  \end{figure}

Next we describe  the shape of the functions $\mu \mapsto F_A(\mu)$ and
$\mu \mapsto F_B(\mu)$
for $\mu \geq 0$,  as illustrated by Figure~\ref{fig:angular}.
This permits us to see the main features of the non-negative
eigenvalues (\ref{5.1}) of $M^U$
furnished by the solutions of (\ref{5.2}) and (\ref{5.3}).


The function $F_A$ (\ref{5.2}) is smooth for $\mu \geq 0$,  except at
the values
\begin{equation}
\mu_m^\infty  = (\nu +1) + 2 m,
\qquad
m=0,1,2,\ldots,
\label{5.14}\end{equation}
where it becomes infinite. More precisely,  as is easily checked
using  the
properties of the $\Gamma$-function, $F_A(\mu)$  approaches
$+ \infty$ and $-\infty$
 as $\mu$ approaches  $\mu_m^\infty$
from above and from below, respectively.
It is positive at $\mu=0$,
and it takes the value zero at
\begin{equation}
\mu_m^0 = (2-\nu) + 2m,
\qquad
m=0,1,2,\ldots
\label{5.15}\end{equation}
Note that
\begin{equation}
\mu_m^0 < \mu^\infty_m < \mu_{m+1}^0 < \mu_{m+1}^\infty,
\qquad
m=0,1,2,\ldots
\label{5.16}\end{equation}
It can be shown (see Appendix C)  that {\em $F_A$ is strictly
monotonically decreasing for
 $\mu \in [0, \mu_0^\infty)$ as well as for $\mu \in
(\mu_m^\infty, \mu_{m+1}^\infty)$ for any $m\geq 0$.}

Since the constant on the right hand side of (\ref{5.2}) is finite,
we conclude from the above that
(\ref{5.2}) admits  {\em a unique solution in the range}
$(\mu_m^\infty, \mu_{m+1}^\infty)$  (\ref{5.14}) for any $m\geq 0$.
There is an additional solution in the range $[0, \mu_0^\infty)$ if
  \begin{equation}
 F_A(0)= \frac{\vert \Gamma(\frac{1+\nu }{2})\vert^2}
{ \vert \Gamma(\frac{2-\nu}{2})\vert^2 } \geq
\frac{\Gamma(\nu +\frac{1}{2})}{\Gamma(-\nu +\frac{3}{2})}
\tan \frac{\alpha \pm \beta}{2}.
\label{5.17}\end{equation}

 Let us sketch the analogous description of the function $F_B(\mu)$
for $\mu\geq 0$.
 As is readily verified,  $F_B(\mu)$
  changes sign from positive to negative
 as $\mu$ passes through the zero locations given by
 \begin{equation}
\bar \mu_0^0= \vert 1-\nu\vert,
\quad
\bar \mu_m^0 = (1-\nu) + 2m, \qquad
m=1,2,\ldots.
\label{5.18}\end{equation}
It becomes $+\infty$ and $-\infty$ as it approaches
\begin{equation}
\bar \mu_m^\infty = \nu + 2m,
\qquad
m=0,1,2, \ldots
\label{5.19}\end{equation}
from above and from below, respectively.
We have
\begin{equation}
\bar \mu_m^0 < \bar\mu_m^\infty < \bar\mu_{m+1}^0 < \bar\mu_{m+1}^\infty,
\qquad
m=0,1,2,\ldots.
\label{5.20}\end{equation}
 Similarly to  the case of $F_A$,
 one can prove that
 {\em $F_B$ is strictly monotonically decreasing   for
$\mu \in [0, \bar \mu_0^\infty)$ as well as
 for $\mu \in (\bar\mu_m^\infty,\bar\mu_{m+1}^\infty)$ for any $m\geq 0$.}

As a consequence of the shape of $F_B$,
for any finite  constant on the right hand side,
(\ref{5.3}) has  {\em a unique solution in the range}
$(\bar \mu_m^\infty, \bar \mu_{m+1}^\infty)$ (\ref{5.19})  for any $m\geq 0$.
There is an additional solution in the range $[0, \bar  \mu_0^\infty)$ if
\begin{equation}
\ F_B(0)= \frac{\vert \Gamma(\frac{\nu }{2})\vert^2 }
{ \vert \Gamma(\frac{1-\nu}{2})\vert^2 } \geq
\frac{\Gamma(\nu +\frac{1}{2})}{\Gamma(-\nu +\frac{3}{2})}
\tan \frac{\alpha \pm \beta}{2} .
\label{5.21}\end{equation}

\subsection{Type 2 eigenvalues}

The formula of the function $F_2$ in (\ref{5.4}) is rather complicated for
general $U$,  for this reason we shall be content with
some remarks on the generic properties of the type 2 eigenvalues.

As an illustration, let us first investigate the equation of type 2
non-positive eigenvalues,
\begin{equation}
F_2(\ri x)= \pm \frac{1}{2},
\qquad x\geq 0,
\label{*5.21}\end{equation}
in the special case $\alpha:= - \frac{\pi}{2}$.
Under this  assumption  $F_2(\ri x)$ simplifies as
\begin{equation}
F_2(\ri x) =
\frac{1}{\sin \beta}\left(   \frac{\cosh \pi x}{\vert \cos  \pi \nu\vert}
+ \frac{\cos\beta}{(6\nu -3)} \left[ a_1 b_1 + a_2 b_2\right](\ri x) \right).
\label{*5.22}\end{equation}
We see from (\ref{4.65}) that $a_k(\ri x) b_k (\ri x) >0$ for $k=1,2$,
and therefore
\begin{equation}
F_2(\ri x) \geq \frac{1}{\sin \beta \vert\cos \pi \nu\vert} >1
\qquad
\hbox{if}\qquad
 0<\beta\leq \frac{\pi}{2}=-\alpha.
\label{*5.23}\end{equation}
Thus $M^U$ does not admit non-positive type 2
eigenvalues for these choices of the
parameters, which include the explicitly solvable case $\beta = -\alpha =
\frac{\pi}{2}$  for which $U=\sigma_1$.
On the contrary, if $\beta > \frac{\pi}{2}=-\alpha$ and
$\beta$ is near enough to $\frac{\pi}{2}$,
then we have
\begin{equation}
F_2(0)>1
\qquad\hbox{and}\qquad
\lim_{x\to \infty} F_2(\ri x) = -\infty,
\label{*5.24}\end{equation}
which implies the existence of at least 2  negative eigenvalues of $M^U$
for the corresponding $U$ (\ref{3.18}).
The second relation in (\ref{*5.24}) holds since for large $x$
the term $a_2(\ri x) b_2(\ri x)$ dominates $F_2$, unless it is multiplied by
zero, as is easily seen from (\ref{5.12}).

The above example shows that, like in the type 1 case, the existence or
non-existence of type 2 negative
eigenvalues depends on the choice of the parameters $\alpha, \beta$.
One can check that
$\lim_{x\to \infty} \vert F_2(\ri x)\vert =\infty$ always holds,
and hence any choice leading
to  $\vert F_2(0)\vert <\frac{1}{2}$ guarantees the
existence of such eigenvalues.
It is also clear that there can be only
finitely many negative eigenvalues,
the maximal number we found in numerical examples is four.

\begin{figure}[ht]  \centering  
\resizebox{.35\columnwidth}{!}{\includegraphics{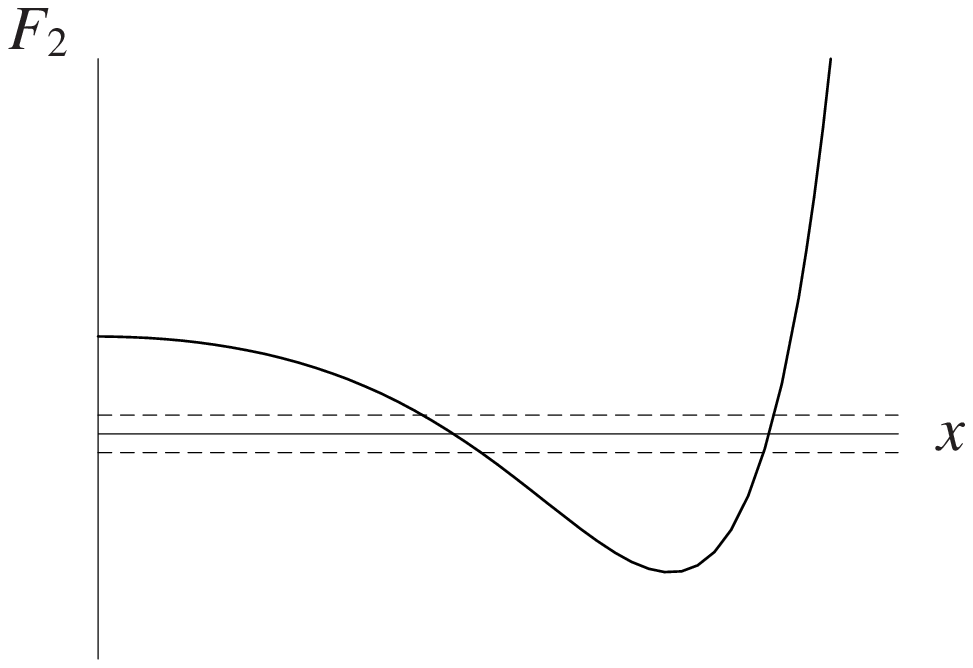}}
     \hskip .15\columnwidth
\resizebox{.35\columnwidth}{!}{\includegraphics{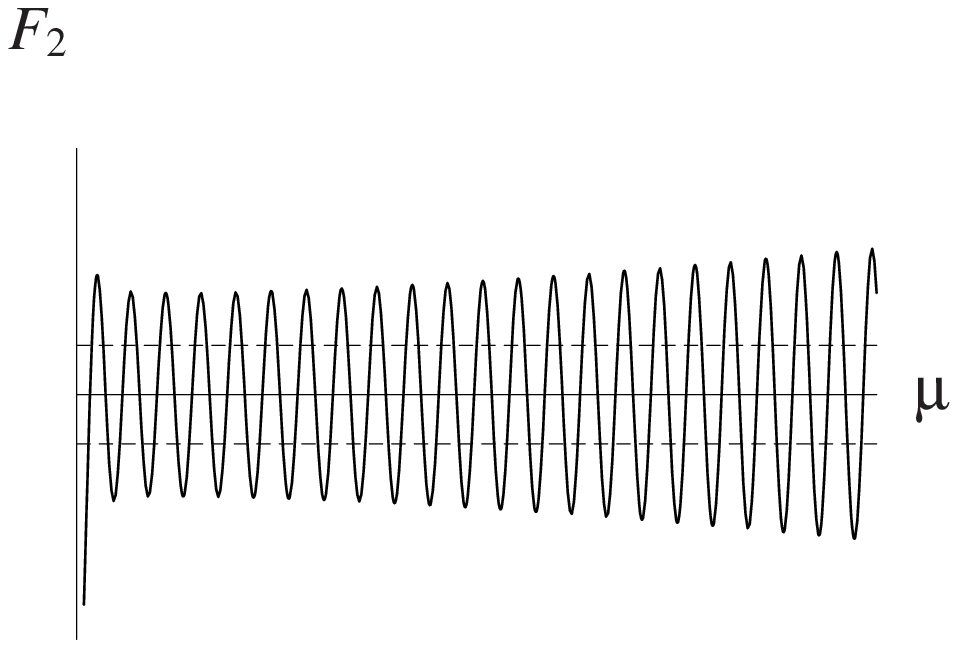}}
\caption{$F_2(\mu)$, for imaginary values of $\mu=\ri x$ (left)
and for real values of $\mu$ (right).
The dashed lines lie at $\pm \frac{1}{2}$. For the
left figure, $\nu = \frac{21}{20}$, $\alpha = \frac{11}{20} \pi$ and
$\beta
= \frac{2}{20} \pi$, a case when the number of solutions of (\ref{5.4}) is
four. For the right figure, the parameters
$\nu = \frac{1001}{1000}$, $\alpha = \frac{3}{10} \pi$, and
$\beta = \frac{715}{1000} \pi$ are chosen so that the plot
exhibits the behaviour of each of the three terms of $F_2$ in (\ref{5.4}).
Initially, the rapidly decreasing middle term dominates, then the
constant amplitude first term rules, and for larger values the third
term
gradually becomes the most significant one. Each term oscillates with
the same
frequency.
}
\label{fig:typetwo}  \end{figure}

\medskip

Since the multiplicity of the $2$-dimensional representations
of $D_6$ in $L^2(S^1)$ is obviously infinite,
$M^U$ has infinitely many type 2 positive eigenvalues.
We can understand the approximate behaviour of
the solutions of (\ref{5.4}) for large
positive $\mu$ by using  the relation (for any real $y$)
\begin{equation}
\lim_{\mu \to \infty} \Gamma(y -\mu) \Gamma (y +\mu)
(y + \mu)^{1-2y} \sin\pi(\mu + 1 -y) = \pi.
\end{equation}
Since $\nu > \frac{1}{2}$, this implies  by (\ref{4.65}) that
$a_1(\mu) b_1(\mu)$ tends to zero as $\mu \to \infty$, and
$a_2(\mu) b_2(\mu)$ asymptotically equals to a constant multiple
of the function
\begin{equation}
f_2(\mu):= (1-\nu + \mu)^{2\nu -1} \sin \pi(\mu + \nu).
\end{equation}
If $(\cos \alpha + \cos \beta)\neq 0$, we thus obtain from
(\ref{5.4}) that $F_2(\mu)$ can be  approximated by a constant
multiple of $f_2(\mu)$ as $\mu\to \infty$.
Therefore, for large enough $\mu$,
 there exist two solutions of $\vert F_2(\mu)\vert =  \frac{1}{2}$
between any two consecutive extrema of the function $f_2(\mu)$,
and actually these solutions lie near to the zeroes of $f_2$.
If $(\cos \alpha + \cos\beta)=0$, then the first term of $F_2$ (\ref{5.4})
is the dominant one. If $\cos \alpha = \cos \beta =0$, then only
this term remains, and we can find the eigenvalues  explicitly as discussed
in Section 7.
If desired, one could derive more precise asymptotic estimates
for the solutions of (\ref{5.4}) by developing the above arguments further.
The behaviour of $F_2$ near to  $\mu=0$ can be rather
complicated in general (see Figure \ref{fig:typetwo}).

\subsection{On the lower-boundedness of the energy as a condition on $U$}

We have seen that  $M^U$ admits, in addition to its infinitely
many non-negative eigenvalues,  a finite number of negative eigenvalues, too,
for certain values of the parameters $\alpha, \beta$ of $U$ (\ref{3.18}).
This result is important, since --- as we shall demonstrate in Section 6 ---
the existence of a negative eigenvalue of $M^U$ implies that the
energy spectrum of the model
(i.e., the spectrum of ${\hat H}_{rel}$ (\ref{2.3}))
 is not bounded from below.
These cases are to be excluded in hypothetical physical applications.
It is complicated to precisely control the conditions for the non-existence of
negative eigenvalues of $M^U$,
but there are certainly cases for which this occurs.
For example, if $\alpha= -\frac{\pi}{2}$ and $0<\beta <\frac{\pi}{2}$,
then $\tan\frac{\alpha \pm \beta}{2} < 0$, and thus neither the
inequalities (\ref{5.6}),  (\ref{5.7}) nor the corresponding equalities
can be satisfied. Together with (\ref{*5.23}) this implies that
the spectrum of $M^U$ is positive for these choices of the
parameters of $U$ (\ref{3.18}).
The same holds, for instance,  for the explicitly
solvable cases $U=\pm {\bf 1}_2, \pm \sigma_1$.

Next, we present a stability result
concerning  the {\em positivity} of the spectrum of $M^U$.
Suppose that all eigenvalues of $M^U$ are positive for some
$U=U(\alpha_0,\beta_0)$ and the parameters $(\alpha_0, \beta_0)$
are generic in the sense that they satisfy the following
inequalities:
\begin{equation}
\vert \tan\frac{\alpha_0 \pm \beta_0}{2}\vert < \infty,
\quad
\sin \beta_0 \neq 0,
\quad
(\cos \alpha_0 + \cos \beta_0)\neq 0.
\label{5.25*}\end{equation}
Then it can be proven that
{\em $M^U$ with $U(\alpha,\beta)$ (\ref{3.18}) has only positive
eigenvalues for any parameters
$(\alpha, \beta)$ near enough to  $(\alpha_0, \beta_0)$.}

To verify the above statement, we first observe that neither any of the
inequalities in (\ref{5.6}),
(\ref{5.7})  nor the corresponding equalities
can hold for $(\alpha,\beta)$ near to
$(\alpha_0, \beta_0)$ by continuity.
 Therefore all type 1 eigenvalues must be positive  for such parameters.
To exclude the possibility of type 2 non-positive eigenvalues,
which would arise from the solutions of (\ref{*5.21}),
notice that $F_2$ (\ref{5.4}) can be written in the form
\begin{equation}
F_2(\ri x, \alpha,\beta)=
e^{\pi x} \left[ \kappa_0(\alpha,\beta) K_0(x) +
\kappa_-(\alpha,\beta) x^{1-2\nu}  K_{-}(x) +
\kappa_+(\alpha,\beta)  x^{2\nu -1} K_+(x) \right],
\label{S.2}\end{equation}
where the functions $K_a$  satisfy
\begin{equation}
K_a(x)>0\quad \forall x\geq 0,\qquad
\lim_{x\to \infty} K_a(x) =1
\qquad (\forall a\in\{0, \pm\}).
\label{S.3}\end{equation}
We see from (\ref{4.65}) and (\ref{5.12}) that the above
relations are valid with
\begin{equation}
K_0(x)= 2 e^{-\pi x} \cosh\pi x,
\quad
K_-(x) = \frac{2\pi  x^{2\nu -1 }
e^{-\pi x}}{\vert \Gamma(\nu + \ri x)\vert^2  },
\quad
K_+(x) = \frac{2\pi  x^{1-2\nu }
e^{-\pi x}}{\vert \Gamma(1-\nu + \ri x)\vert^2  }
\label{S.4}\end{equation}
and
\begin{eqnarray}
&&\kappa_0(\alpha,\beta) = \frac{1}{2 \cos \pi \nu}
\frac{\sin \alpha}{\sin \beta}, \nonumber\\
&&\kappa_- (\alpha,\beta)= \frac{\Gamma^2(\nu+\frac{1}{2})
2^{2(\nu -1)}}{\pi (2\nu-1)}
\frac{\cos \beta - \cos \alpha}{\sin \beta},
\nonumber\\
&&\kappa_+(\alpha,\beta)= \frac{\Gamma^2(\frac{3}{2}-\nu )
2^{-2\nu }}{\pi (2\nu-1)}
\frac{\cos \beta + \cos \alpha}{\sin \beta}.
\label{S.5}\end{eqnarray}
Let us assume that (\ref{5.25*}) is satisfied and
$\kappa_+(\alpha_0, \beta_0) >0$ (the case of the other sign is similar).
We can choose a neighbourhood $V_1$ of $(\alpha_0, \beta_0)$ and
constants $\gamma_a >0$ so that
\begin{equation}
\kappa_+(\alpha, \beta) > \gamma_+,
\quad
\vert \kappa_0(\alpha, \beta)\vert < \gamma_0,
\quad
\vert \kappa_-(\alpha, \beta)\vert < \gamma_-,
\qquad \forall (\alpha,\beta) \in V_1.
\label{S.6}\end{equation}
Then we fix some $x_0>0$ for which
\begin{equation}
F_2(\ri x, \alpha,\beta) \geq e^{\pi x} \left[
\gamma_+  x^{2\nu -1} K_+(x)
-\gamma_0 K_0(x)
- \gamma_- x^{1-2\nu}  K_{-}(x)\right] > 1,
\qquad \forall  x>x_0
\label{S.7}\end{equation}
and $\forall (\alpha, \beta) \in V_1$.
This ensures that there is no solution of (\ref{*5.21}) for $x> x_0$.
Since  $K_0(x)$ and $x^{\pm (2\nu -1)} K_\pm (x)$ are bounded on
$[0, x_0]$,
for any $\epsilon >0$ we can find
a neighbourhood $V_2$ of $(\alpha_0, \beta_0)$ so that
\begin{equation}
\vert F_2(\ri x, \alpha,\beta) - F_2(\ri x, \alpha_0, \beta_0)\vert
 \leq \epsilon \qquad\hbox{if}\qquad
(\alpha, \beta) \in V_2, \quad 0\leq x\leq x_0.
\end{equation}
By choosing $\epsilon$ appropriately, for instance in such a way that
\begin{equation}
F_2(x, \alpha_0, \beta_0) - \frac{1}{2} \geq 2\epsilon
\quad
\hbox{for}\quad 0\leq x\leq x_0,
\end{equation}
we conclude that (\ref{*5.21}) has no solution
if $(\alpha,\beta) \in V_1 \cap V_2$.
This implies the stability result that we wanted to prove.

We can establish  a counterpart of the  above stability result
concerning the  `impermissible' boundary conditions as well,
for which negative eigenvalues of the angular Hamiltonian  exist.
Namely, if
$M^U$ has one or more negative eigenvalues, then generically this
property is stable under arbitrary small perturbations of
$U$ in (\ref{3.18}).
Indeed, this is the case obviously if $M^U$ admits a type 1
negative eigenvalue or a type 2 negative eigenvalue
which is generic in the sense
that it arises from  the graph of $x\mapsto
\vert F_2(\ri x)\vert$
properly intersecting, not just touching,  the horizontal line
located at $\frac{1}{2}$.

\section{The radial Hamiltonian}
\setcounter{equation}{0}

Recall from (\ref{1.7}) that, at the formal level,  the radial
Hamiltonian   reads
 \begin{equation}
H_{r,\lambda} = - \frac{d^2}{d r^2} - \frac{1}{r} \frac{d}{d r}  +
\frac{3}{8} \omega^2 r^2 +
\frac{\lambda}{r^2}.
\label{6.1}\end{equation}
After having characterized  the qualitative features of the eigenvalues
$\lambda = (3 \mu)^2$  of $M^U$,
we below analyze  the energy levels of the relative motion of the
three particle Calogero system defined by the eigenvalues of the
possible self-adjoint versions of
the radial Hamiltonian.

Since $H_{r,\lambda}$ has to be self-adjoint on a domain in
$L^2({\bf R}_+, r dr)$,
it is more convenient to deal with the equivalent operator
\begin{equation}
\cH_{r,\lambda} := \sqrt{r} \circ H_{r,\lambda} \circ \frac{1}{\sqrt{r}} =
- \frac{d^2}{d   r^2} +\frac{3}{8} \omega^2 r^2 +
\frac{\lambda-\frac{1}{4}}{r^2},
\label{6.2}\end{equation}
which must be self-adjoint on a corresponding domain in  $L^2({\bf R}_+, dr)$.
It is easy to check that,  for any eigenvalue, both
of the two independent eigenfunctions of
the differential operator $\H_{r,\lambda}$ are
locally square integrable around $r=0$ if and only if $\lambda <1$.
For this reason \cite{DS,Richt,Meetz},
$\cH_{r, \lambda}$ admits inequivalent
choices of self-adjoint domains
 if and only if $\lambda < 1$.
It follows from general theorems, collected in Appendix A from \cite{DS},
that any self-adjoint version of  $\cH_{r, \lambda}$ possesses {\em pure
discrete spectrum}.

If $\lambda \geq 1$, then the unique self-adjoint domain of
$\cH_{r,\lambda}$ consists of those
complex functions
$\rho$ on ${\bf R}_+$ for which $\rho$ and $\rho'$ are absolutely
continuous away from $r=0$
and both $\rho$ and $\cH_{r,\lambda} \rho$ belong to $L^2({\bf R}_+, dr)$.
It is straightforward to show that the spectrum is given by the eigenvalues
\begin{equation}
E_{m,\lambda}=  2 c ( 2m +1+ \sqrt{\lambda} ),
\quad c:=  \sqrt\frac{3}{8} \omega,
 \quad m=0,1,2,\ldots,
\label{6.3}\end{equation}
with  the corresponding  eigenfunctions
\begin{equation}
\rho_{m,\lambda}(r) = r^{\frac{1}{2} + \sqrt{\lambda} }
e^{-\frac{1}{2} c  r^2} L_m^{\sqrt{\lambda}}(c r^2),
\label{6.4}\end{equation}
where $L_m^{\sqrt{\lambda}}$ is the (generalized)
Laguerre polynomial \cite{GR}, $\sqrt{\lambda}\geq 1$.
This result  is  contained, for example,  in  \cite{Cal3,Landau}.

 From now on  we consider the case
\begin{equation}
\lambda <1.
\label{6.5}\end{equation}
Let $\varphi_1$ and $\varphi_2$ be two independent real
eigenfunctions of $\cH_{r,\lambda}$
associated with an arbitrary real eigenvalue.
In fact (see  \cite{Richt,Krall}),
in addition to the same properties they have for $\lambda \geq 1$,
the functions $\rho$ in a self-adjoint domain of $\cH_{r,\lambda}$  must now
also satisfy a {\em boundary condition} of the following form:
\begin{equation}
\frac{ W[\rho, \varphi_1]_{0+}}
{W[\rho, \varphi_2]_{0+}}=\kappa(\lambda),
\label{6.6}\end{equation}
where $\kappa(\lambda)$ is a real number or is infinity.
Here $\kappa(\lambda)=0$ means that $W[\rho,\varphi_1]_{0+}=0$, and
similarly $W[\rho,\varphi_2]_{0+}=0$ is required if $\kappa(\lambda)$
is infinite.
Our notation emphasizes that one can in principle  choose different  constants
on the right hand  side of (\ref{6.6}) for different $\lambda$.
We remark that the condition (\ref{6.6}) can be regarded as a special
case of the boundary conditions of the form in (\ref{3.14}),
where one restricts to the positive side of the singular point
$r=0$
of $\cH_{r,\lambda}$ {\em considered on} $\bR$
(accordingly $U$ reduces to a phase), prohibiting the particle
from going into $r<0$ from $r>0$.
With the `reference modes'  $\varphi_k$ fixed subsequently,
the self-adjoint radial Hamiltonian specified by condition (\ref{6.6})  is
denoted as  $\cH_{r,\lambda, \kappa(\lambda)}$.
(In the notation used in (\ref{2.3}),
one may substitute
 ${\hat H}_{r,\lambda,\kappa(\lambda)}: =
\frac{1}{\sqrt{r}}\circ  \cH_{r,\lambda,\kappa(\lambda)} \circ \sqrt{r}$
 for ${\hat H}_{r,\lambda}$.)

 In order to determine the spectrum of  $\cH_{r,\lambda, \kappa(\lambda)}$,
 we first write down the solutions of
 \begin{equation}
\H_{r,\lambda} \rho = E \rho
\label{6.7}\end{equation}
 for any real number $E$.
 To do this, it is convenient to introduce
 \begin{equation}
\sigma:= c r^2,
\qquad
\xi:= \frac{E}{4c} - \frac{\sqrt{\lambda} +1}{2}
\label{6.8}\end{equation}
with $c$ given in (\ref{6.3}).
Then one can check that, if $\lambda\neq 0$,   two
{\em independent}\footnote{To save space,  we
henceforth exclude the $\lambda =0$ case from our investigation,
since it would require a separate treatment and the final result
for $\lambda=0$ is
expected to be similar to that for any $0<\lambda<1$. }
 solutions of (\ref{6.7}) are provided by the functions
\begin{eqnarray}
&&\rho_{E,1}(r) = \sigma^{\frac{1}{2}(\frac{1}{2} + \sqrt{\lambda})}
e^{-\frac{1}{2}\sigma} \Phi(-\xi, \sqrt{\lambda} +1, \sigma),
\nonumber\\
&&
 \rho_{E,2}(r) = \sigma^{\frac{1}{2}(\frac{1}{2} - \sqrt{\lambda})}
 e^{-\frac{1}{2}\sigma} \Phi(-\xi- \sqrt{\lambda}, 1-\sqrt{\lambda}, \sigma),
\label{6.9}\end{eqnarray}
where $\Phi(a,b,z)$ is the  confluent hypergeometric function,
also called Kummer's function.
Here, as before, we insist on the convention that either
$\sqrt{\lambda}=3\mu\geq 0$ or
its imaginary part is positive.
Up to a multiplicative factor,  there is a unique  linear combination
of the functions in (\ref{6.9})
that lies in  $L^2({\bf R}_+, dr)$,  given by
\begin{equation}
\rho_E = \frac{\Gamma(1-\sqrt{\lambda})}{\Gamma(-\xi - \sqrt{\lambda})}
\rho_{E,1} - \frac{\Gamma(1+\sqrt{\lambda})}{\Gamma(-\xi )} \rho_{E,2}.
\label{6.10}\end{equation}
In fact, with an irrelevant factor $C(\sqrt{\lambda})$, one has
\begin{equation}
\rho_E(r)= C(\sqrt{\lambda}) \sigma^{\frac{1}{2}(\frac{1}{2} + \sqrt{\lambda})}
e^{-\frac{1}{2}\sigma} U(-\xi, 1+\sqrt{\lambda}, \sigma),
\label{6.11}\end{equation}
where the function $U$ satisfies
$U(\xi, 1+ \sqrt{\lambda},\sigma)= \sigma^\xi[ 1 + O(\sigma^{-1})]$
as $\sigma \to \infty$ (see eq.~13.1.8 in \cite{AS}).
This guarantees the square integrability of $\rho_E$.
For $E$ to belong to the spectrum of  $\cH_{r,\lambda,\kappa(\lambda)}$,
$\rho_E$ must satisfy the boundary condition (\ref{6.6}).

\begin{figure}[ht]  \centering  
\resizebox{.45\columnwidth}{!}{\includegraphics{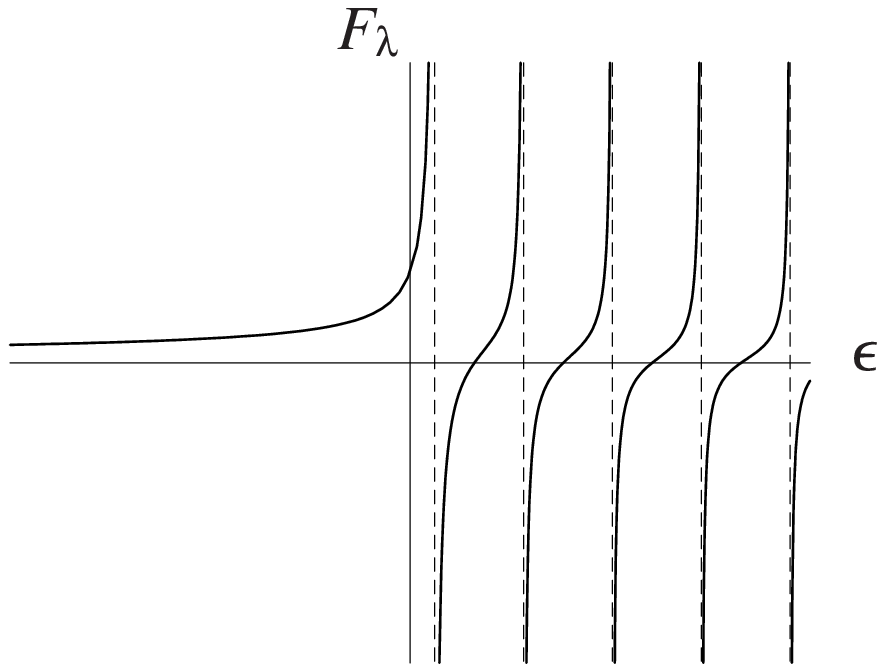}}
\caption{$F_\lambda$, as the function of $\epsilon$ [see (\ref{6.16})],
for $\lambda = 1/5$. The dashed lines, where the function diverges, are
located at $\epsilon_m^\infty$ (\ref{6.18}).}
\label{fig:radial}  \end{figure}

Let us now suppose that
\begin{equation}
0 < \lambda < 1.
\label{6.12}\end{equation}
In this  case the $\rho_{E,k}$ are real functions, and we
fix our reference modes to be
\begin{equation}
\varphi_k:= \rho_{E_0,k}
\label{6.13}\end{equation}
with some arbitrary real $E_0$.
For arbitrary $E$ and $E_0$,    an easy calculation yields
\begin{equation}
W[\rho_{E,k}, \rho_{E_0,l}]_{0+}:=
\lim_{r\to 0} \left( \rho_{E,k} \frac{d \rho_{E_0,l}}{dr} -
 \rho_{E_0,l} \frac{d \rho_{E,k}}{dr}\right)(r)
  = - 2  \epsilon_{k,l} \sqrt{\lambda} \sqrt{c},
\label{6.14}\end{equation}
where  $\epsilon_{k,l}$ is the usual alternating tensor.
Therefore we find that
\begin{equation}
\frac{ W[\rho_E, \varphi_1]_{0+}}
{W[\rho_E , \varphi_2]_{0+}}=
\frac{ \Gamma(1+\sqrt{\lambda} ) \Gamma( -\xi - \sqrt{\lambda})}
 {\Gamma(1-\sqrt{\lambda} ) \Gamma( -\xi )}.
\label{6.15}\end{equation}
By substituting back $\xi$ (\ref{6.8}),  we obtain the
following condition that {\em determines
the eigenvalues of $\cH_{r,\lambda, \kappa(\lambda)}$ under (\ref{6.12})}:
\begin{equation}
F_\lambda(\epsilon):= \frac{  \Gamma( -\epsilon +\frac{1- \sqrt{\lambda}}{2})}
{\Gamma( -\epsilon +\frac{1+ \sqrt{\lambda}}{2})}=
- \frac{\Gamma(-\sqrt{\lambda})}{\Gamma(\sqrt{\lambda})}  \kappa(\lambda)
\qquad\hbox{with}\qquad
\epsilon:= \frac{E}{4c}.
\label{6.16}\end{equation}
Next, we  analyze the shape of the function
$F_\lambda$, which is illustrated by Figure~\ref{fig:radial}.

Let us start  by observing  that $F_\lambda$ has zeroes at
\begin{equation}
\epsilon_m^0 =\frac{1 +\sqrt{\lambda}}{2} +m,
\qquad
\forall m=0,1,2,\ldots
\label{6.17}\end{equation}
and it becomes $\pm \infty$ at $\epsilon_m^\infty \mp 0$ for
\begin{equation}
\epsilon_m^\infty =\frac{1 -\sqrt{\lambda}}{2} +m,
\qquad
\forall m=0,1,2,\ldots
\label{6.18}\end{equation}
By considering
\begin{equation}
\frac{F_\lambda'(\epsilon)}{F_\lambda(\epsilon)} =
\psi(-\epsilon+ \frac{1 +\sqrt{\lambda}}{2} )
- \psi(-\epsilon +\frac{1 -\sqrt{\lambda}}{2})
\label{6.19}\end{equation}
with the notation (\ref{5.9}),  wee see immediately that for
\begin{equation}
-\infty < \epsilon < \epsilon_0^\infty
\label{6.20}\end{equation}
$F_\lambda(\epsilon)>0$ and $F_\lambda'(\epsilon)>0$.
This and the asymptotics of $F_\lambda$ imply  that
$F_\lambda(\epsilon)$ monotonically
increases from $0$ to $+\infty$ as $\epsilon$
varies from $-\infty$ to $\epsilon_0^\infty$.
If $\epsilon > \epsilon_0^\infty$,  we use the
reflection formula  (\ref{D.2}) for $\psi$ to write
\begin{equation}
\frac{F_\lambda'(\epsilon)}{F_\lambda(\epsilon)} =
\left[\psi(\epsilon+ \frac{1 -\sqrt{\lambda}}{2} )
- \psi(\epsilon +\frac{1 +\sqrt{\lambda}}{2})\right]
+
\pi\left[\cot \pi(\epsilon+ \frac{1 -\sqrt{\lambda}}{2} )
- \cot\pi (\epsilon +\frac{1 +\sqrt{\lambda}}{2})\right].
\label{6.21}\end{equation}
We have
\begin{equation}
\epsilon_m^\infty  < \epsilon_m^0 < \epsilon_{m+1}^\infty,
\qquad \forall m=0,1,2,\ldots
\label{6.22}\end{equation}
If
\begin{equation}
\epsilon_m^\infty < \epsilon < \epsilon_m^0,
\label{6.23}\end{equation}
 then both differences in the square brackets in (\ref{6.21}) are negative.
Since under (\ref{6.23}) $F_\lambda(\epsilon)<0$,  it follows that
$F_\lambda$ is monotonically increasing in this domain.
If
\begin{equation}
\epsilon_m^0 < \epsilon < \epsilon_{m+1}^\infty,
\label{6.24}\end{equation}
then $F_\lambda(\epsilon)>0$ and  the differences in (\ref{6.21})
have opposite signs.
We can show that $F_\lambda'(\epsilon)>0$ by using the integral
formulae (\ref{D.12}) and (\ref{D.13}) in the same way as in Appendix C,
which proves that $F_\lambda$ is increasing in this domain as well.

Supposing that   $\kappa(\lambda)  \notin\{0, \infty\}$,
it is clear from the shape of the function $F_\lambda$ that
there exists precisely one positive
eigenvalue, $\frac{E}{4c}$,   in each interval
$(\epsilon_m^\infty, \epsilon_{m+1}^\infty)$ for any
non-negative integer $m$.
Moreover, one obtains at most one negative eigenvalue, which occurs
precisely if
\begin{equation}
F_\lambda(0)= \frac{  \Gamma( \frac{1- \sqrt{\lambda}}{2})}
{\Gamma( \frac{1+ \sqrt{\lambda}}{2})} >
-  \frac{\Gamma(-\sqrt{\lambda})}{\Gamma(\sqrt{\lambda})} \kappa(\lambda) >0.
\label{6.25}\end{equation}
There is also a non-negative eigenvalue in $[0, \epsilon_0^\infty)$ if
$F_\lambda(0) \leq -
\frac{\Gamma(-\sqrt{\lambda})}{\Gamma(\sqrt{\lambda})} \kappa(\lambda) $.

The eigenvalue equation (\ref{6.16})
is explicitly solvable if $\kappa(\lambda)=0$ or
$\kappa(\lambda)=\infty$.
In the former case the eigenvalues have the same form as in
(\ref{6.3}), and the
corresponding eigenfunction reduces to $\rho_{E,1}$ in (\ref{6.9}),
which has the same form
as (\ref{6.4}) up to an irrelevant constant.
If $\kappa(\lambda)=\infty$, then we find the eigenvalues
\begin{equation}
\tilde E_{m,\lambda} = 2c (2m + 1 - \sqrt{\lambda}),
\qquad
m=0,1,2,\ldots
\label{6.26}\end{equation}
Under $E=\tilde E_{m,\lambda}$,  $\rho_E$ (\ref{6.10}) is
proportional to $\rho_{E,2}$,  which (up to another irrelevant factor)
gives the eigenfunction
\begin{equation}
\tilde \rho_{m,\lambda}(r) = r^{\frac{1}{2} - \sqrt{\lambda} }
e^{-\frac{1}{2} c  r^2} L_m^{-\sqrt{\lambda}}(c r^2),
\label{6.27}\end{equation}
Note that  all energy levels  are positive in these cases.

Our spectral condition (\ref{6.16}) is consistent with the result
in \cite{indPLA}\footnote{In \cite{indPLA}  the shape
of $F_\lambda$ in (\ref{6.16}) was illustrated by a Mathematica plot,
without presenting a proof of its properties as supplied above.},
where
the inequivalent quantizations of the radial Hamiltonian $H_{r,\lambda}$
(\ref{1.7})
 were considered by using
a different method,
  with $0<\lambda<1$ taken from the
eigenvalues of two special self-adjoint
versions of the angular Hamiltonian $H_\Omega$ (\ref{1.6})
that are well-understood for any $N$ due to Calogero \cite{CalN}.
In our $N=3$ case-study  those correspond to $M^U$ with $U=-{\bf 1}_2$ or
$U={\bf 1}_2$.

Let us now deal with the case when
\begin{equation}
\sqrt{\lambda} =3\mu =  \ri x
\quad\hbox{with some}\quad
x>0.
\label{6.28}\end{equation}
Then the functions $\rho_{E,k}$ are complex, and their real and
imaginary parts are also
solutions of (\ref{6.7}).
We choose our reference modes to be (up to a factor) the
real and imaginary parts
of $\rho_{E_0,1}$ for some real $E_0$.
Explicitly, with $\xi_0= \frac{E_0}{4c} - \frac{\ri x +1}{2}$, we define
\begin{eqnarray}
&&\varphi_1(r):= r^{\frac{1}{2}} e^{-\frac{1}{2}\sigma}
\left[\Re \Phi(-\xi_0, 1+\ri x,\sigma)   \cos (x\log r)
-\Im \Phi(-\xi_0, 1+\ri x,\sigma)   \sin (x\log r)\right]
\qquad \quad\nonumber\\
&&\varphi_2(r):= r^{\frac{1}{2}}
e^{-\frac{1}{2}\sigma}\left[\Re \Phi(-\xi_0, 1+\ri x,\sigma)   \sin (x\log r)
+\Im \Phi(-\xi_0, 1+\ri x,\sigma)   \cos (x\log r)\right].
\label{6.29}\end{eqnarray}
These eigenfunctions of $\H_{r, \lambda}$ are independent, since we find
\begin{equation}
W[\varphi_1, \varphi_2](r)= \lim_{r\rightarrow 0+} W[\varphi_1,\varphi_2](r)=x.
\label{6.30}\end{equation}
One also readily  calculates that
\begin{eqnarray}
&&\lim_{r\rightarrow 0+} W[\rho_{E,1},\varphi_1](r)=
-\ri x c^{\frac{1}{2}( {\frac{1}{2}+\ri x})}
\nonumber\\
&&\lim_{r\rightarrow 0+} W[\rho_{E,2},\varphi_1](r)=
\ri x c^{\frac{1}{2}( {\frac{1}{2}-\ri x})}
\nonumber\\
&&\lim_{r\rightarrow 0+} W[\rho_{E,1},\varphi_2](r)=
x c^{\frac{1}{2}( {\frac{1}{2}+\ri x})}
\nonumber\\
&&\lim_{r\rightarrow 0+} W[\rho_{E,2},\varphi_2](r)=
x c^{\frac{1}{2}( {\frac{1}{2}-\ri x})}.
\label{6.31}\end{eqnarray}
By inserting these into (\ref{6.6}) for $\rho=\rho_E$  (\ref{6.10}), we obtain
the following eigenvalue equation:
\begin{equation}
\cot \arg \left( c^{\frac{1}{2} \ri x} \Gamma(1-\ri x)
\Gamma(-\epsilon + \frac{1 + \ri x}{2})
\right)= - \kappa(\lambda)
\qquad  (\epsilon = \frac{E}{4c}).
\label{6.32}\end{equation}
Equivalently, we have to solve
\begin{equation}
\frac{\Re \left(c^{\frac{1}{2} \ri x} \Gamma(1-\ri x)
\Gamma(-\epsilon + \frac{1 + \ri x}{2}) \right)}
{\Im \left(c^{\frac{1}{2} \ri x} \Gamma(1-\ri x)
\Gamma(-\epsilon + \frac{1 + \ri x}{2})\right) }
= - \kappa(\lambda).
\label{6.33}\end{equation}
Because of the shape of the potential in (\ref{6.2}),
one expects to find infinitely many solutions
for $\epsilon$ around  $+\infty$ as well as around  $-\infty$.

Indeed,
we can easily prove that {\em the accumulation points of the spectrum
of $\H_{r,\lambda, \kappa(\lambda)}$ are precisely $\pm\infty$.}
For this purpose, it is advantageous to rewrite (\ref{6.32}) as
\begin{equation}
 \Omega(\epsilon, x):=
 \arg\Gamma(-\epsilon + \frac{1}{2} + \frac{\ri x}{2}) = \vartheta(x,
 \kappa(\lambda),c) \quad \hbox{mod} \,\, \pi,
\label{6.34}\end{equation}
where
\begin{equation}
\vartheta(x,\kappa(\lambda),c)= \mbox{arccot}
(-\kappa(\lambda)) -\arg\Gamma(1-\ri x) - \frac{x}{2}\log c .
\label{6.35}\end{equation}
In (\ref{6.34}) we can take   $\Omega(\epsilon,x)$ to be the smooth function
of $\epsilon \in \bR$ defined  by
\begin{equation}
 \Omega(\epsilon, x):= \
 \arg\Gamma( \frac{1}{2} + \frac{\ri x}{2}) + \int_0^\epsilon dy\, \omega(y,x),
 \label{6.36}\end{equation}
 where  the ambiguity in $\arg\Gamma( \frac{1}{2} +
\frac{\ri x}{2})$ is fixed arbitrarily and
 \begin{equation}
 \omega(y,x):=\frac{ d}{d y}
\arg\Gamma(-y + \frac{1}{2} + \frac{\ri x}{2})  =
\frac{1}{2\ri } \left[ \psi(-y + \frac{1}{2} - \frac{\ri x}{2})-
\psi(-y  + \frac{1}{2} + \frac{\ri x}{2} )\right].
 \label{6.37}\end{equation}
 It follows by means of (\ref{5.10}) that $\omega(y,x)<0$, and
one sees with the help of
 the asymptotic expansion of $\psi$   and (\ref{6.36})
 (or directly from the asymptotic expansion of $\Gamma$) that
$\lim_{\epsilon \to\pm \infty}  \Omega(\epsilon, x)= \mp \infty$.
Therefore $\Omega(\epsilon,x)$ decreases monotonically from
$+\infty$ to $-\infty$ as
$\epsilon$ runs through the real axis. This implies that the set
of solutions of
(\ref{6.34}) is bounded neither from below nor from above, and
there are finitely many
solutions in any finite interval.

In the foregoing  derivation  we assumed that $\kappa(\lambda)\neq 0$ and
$\kappa(\lambda)\neq  \infty$,
but the conclusion remains valid for these
special  values, too, as one can confirm by inspection of the
respective conditions $W[\rho_E,\varphi_1]_{0+}=0$ and
$W[\rho_E,\varphi_2]_{0+}=0$.

Since  for $\lambda <0$ the energy spectrum is not bounded from below,
in physical applications
one has to  exclude those connection matrices $U$ for
which the angular Hamiltonian $M^U$ possesses a negative  eigenvalue.

\section{Explicitly solvable cases}
\setcounter{equation}{0}

We illustrate our inequivalent quantizations of the $N = 3$ Calogero model
by considering a few special cases which can be solved explicitly.  Recall
that our inequivalent quantizations are specified by the parameter
$\kappa(\lambda)$  (to each $\lambda<1$) for
the radial part (\ref{6.6})
and the parameters in $U$ (\ref{3.18}) for the angular part.
In this section $\lambda$ will always be positive, and
for simplicity we adopt
the choice $\kappa(\lambda) = 0$ for any $0<\lambda<1$.
Then the energy eigenvalues and the radial solutions have
the form given by  (\ref{6.3}) and (\ref{6.4}) {\em for all} $\lambda >0$.
For the angular part, we consider the four
cases, $U = -{\bf 1}_2$, $U = {\bf 1}_2$,
$U = \sigma_1$ and $U = -\sigma_1$.
We shall see, in particular, that the case $U = \sigma_1$ admits a
smooth limit for $\nu \to 1$ (i.e., $g \to 0$) in which the system
defined by ${\hat H}_{rel}$ (\ref{2.3})
becomes the harmonic oscillator in two dimensions.
For $U\neq \sigma_1$ the system resulting in the
$g\to 0$ limit can be interpreted as the harmonic oscillator
plus an extra singular potential in the angular sector,
which is supported at the points of $\cS$ (\ref{3.5})
and manifests itself in the boundary condition.

The spectrum of the angular Hamiltonian for
$U = \pm {\bf 1}_2, \pm \sigma_1$ and the energy
spectrum for $U=\sigma_1$ are summarized by Figure 5
at the end of the section.

\subsection{The case $U = -{\bf 1}_2$}

We begin by the `Dirichlet' case $U = -{\bf 1}_2$,
which
is in fact the standard choice of boundary condition that
has been used since the introduction of the model \cite{Cal3}.
As $U = -{\bf 1}_2$ is one of the separating cases discussed in Section 4.2,
we just recall (\ref{4.21})--(\ref{4.25}) for the eigenstates of the
angular part.  With the eigenvalues, the solutions are
\begin{eqnarray}
&&
\eta_\mu^{A }(\phi) = \sum_{k=1}^6 C_-^k \eta_{-,\mu}^k(\phi),
\qquad
\mu = 2n+1 +\nu ,
\nonumber\\
&&
\eta_\mu^{B }(\phi) = \sum_{k=1}^6 C_+^k \eta_{+,\mu}^k(\phi),
\qquad
\mu = 2n +\nu,
\label{7.1.1}\end{eqnarray}
for $n=0,1,2,\ldots$.  The arbitrary coefficients $C_\pm^k$, $k = 1, 2,
\ldots, 6$ show that all levels have multiplicity 6.

These solutions $\eta_\mu^A$, $\eta_\mu^B$ are then combined with the
solutions for the radial part
$R_{m, \lambda}(r) = r^{-{1\over 2}} \rho_{m,\lambda}(r)$, where
$\rho_{m,\lambda}$ is given in (\ref{6.4}), to form
the eigenstates for the entire system governed by ${\hat H}_{rel}$.  Since
$\lambda = (3\mu)^2$ is determined by $n$, those states  (modulo the
normalization constant) may be presented
as ($m, n = 0, 1, 2, \ldots$)
\begin{eqnarray}
&&
\Psi^{A}_{mn}(r, \phi) =
R_{m, \lambda}(r)\,\eta_\mu^{A}(\phi),
\qquad
E^{A}_{mn} = 2c\left(2m + 1 + 3( 2n +1 + \nu )\right),
\nonumber\\
&&
\Psi^{B}_{mn}(r, \phi) =
R_{m, \lambda}(r)\,\eta_\mu^{B}(\phi),
\qquad
E^{B}_{mn} = 2c\left(2m + 1 + 3( 2n + \nu )\right),
\label{7.1.2}\end{eqnarray}
where $2c = \sqrt{{3\over 2}}\, \omega$; see (\ref{6.3}).

On account of the multiplicity, one can choose the eigenstate in any
particular representation of the exchange-$S_3$ (or more generally the
$D_6$) symmetry group.  For example, if we choose $C_-^k = (-1)^{k-1} C_-^1$
for $k=2,\ldots,6$, then the resultant state $\eta_\mu^{A}$ for
the angular part
becomes a $(+1)$ eigenstate of all the reflections $\hat P_n$
and hence it is bosonic.  On the
other hand, if we choose $C_-^k = C_-^1$, then the resultant
state $\eta_\mu^{A}$ becomes a $(-1)$ eigenstate and hence it is fermionic.
In contrast, for $\eta_\mu^{B}$ the choice
$C_+^k = (-1)^{k-1} C_+^1$ makes it fermionic and $C_+^k = C_+^1$ makes it
bosonic.
At this point we
recall that the basic solutions $v_{i, \mu}(\phi)$, $i = 1, 2$, defined in
(\ref{4.1}) are periodic in $\phi$ with period ${\pi\over 3}$ and are
symmetric with respect to $\phi = {\pi\over 6}$.  Thus, if we
introduce the sign factors,
\begin{equation}
c(\phi) := {{\cos 3\phi}\over{\vert \cos 3\phi\vert}},
\qquad
s(\phi) := {{\sin 3\phi}\over{\vert \sin 3\phi\vert}},
\qquad
t(\phi) := {{\tan 3\phi}\over{\vert \tan 3\phi\vert}},
\label{7.1.3}\end{equation}
we can express these bosonic and fermionic eigenstates concisely over
$S^1\setminus \cS$ in terms of the basic functions.  For instance, the
bosonic states may be presented as
\begin{equation}
\eta_\mu^{A}(\phi) = c(\phi)\, v_{1, \mu}(\phi),
\qquad
\eta_\mu^{B}(\phi) = v_{1, \mu}(\phi),
\label{7.1.4}\end{equation}
for the levels $\mu$ given, respectively, by (\ref{7.1.1}).
It is easily confirmed that the sign factor $c(\phi)$
attached in (\ref{7.1.4}) takes care of the sign conventions required for
$\eta_{-,\mu}^k$ together with the choice of the coefficients needed to
provide the bosonic states.  Similarly, the fermionic states are given by
\begin{equation}
\eta_\mu^{A}(\phi) = t(\phi)\, v_{1, \mu}(\phi),
\qquad
\eta_\mu^{B}(\phi) = s(\phi)\, v_{1, \mu}(\phi).
\label{7.1.5}\end{equation}
For the eigenvalues $\mu$ in (\ref{7.1.1}), one
may use the relation
\begin{equation}
F\left(-{l\over 2},\, {l\over 2}+\nu,\, \nu + {1\over 2};\, \sin^2 3\phi
\right)
= {{l!\, \Gamma(2\nu)}\over{\Gamma(l+2\nu)}}\,
C_l^\nu(\vert \cos 3\phi\vert)
\label{7.1.6}\end{equation}
to replace $v_{1, \mu}$ in (\ref{7.1.4}) and (\ref{7.1.5}) with the
Gegenbauer polynomial $C_l^\nu$.
These bosonic and fermionic eigenstates recover the original solutions
obtained by Calogero for $N=3$ \cite{Cal3}.

\subsection{The case $U = {\bf 1}_2$}

The `Neumann' case $U = {\bf 1}_2$ can be solved analogously to
the preceding `Dirichlet' case;
eq.~(\ref{4.21}) with (\ref{4.26-27}) yields
the angular solutions
\begin{eqnarray}
&&
\eta_\mu^{A }(\phi) = \sum_{k=1}^6 C_-^k \eta_{-,\mu}^k(\phi),
\qquad
\mu = 2n+1 + (1-\nu) ,
\nonumber\\
&&
\eta_\mu^{B }(\phi) = \sum_{k=1}^6 C_+^k \eta_{+,\mu}^k(\phi),
\qquad
\mu = \vert 2n + (1-\nu)\vert,
\label{7.2.1}\end{eqnarray}
for $n=0,1,2,\ldots$, with arbitrary coefficients
$C_\pm^k$, $k = 1, \ldots,6$.
The eigenstates and the eigenvalues for the entire system thus read
\begin{eqnarray}
&&
\Psi^{A}_{mn}(r, \phi) =
R_{m, \lambda}(r)\,\eta_\mu^{A}(\phi),
\qquad
E^{A}_{mn} = 2c\left(2m + 1 + 3( 2n +1 + (1-\nu) )\right),
\nonumber\\
&&
\Psi^{B}_{mn}(r, \phi) =
R_{m, \lambda}(r)\,\eta_\mu^{B}(\phi),
\qquad
E^{B}_{mn} = 2c\left(2m + 1 + 3\vert 2n + (1-\nu)\vert \right).
\label{7.2.2}\end{eqnarray}

As in the previous case, the eigenstates can be made bosonic or fermionic by
choosing the coefficients $C_\pm^k$, appropriately.  Concise forms for these
states are also available as
\begin{equation}
\eta_\mu^{A}(\phi) = c(\phi)\, v_{2, \mu}(\phi),
\qquad
\eta_\mu^{B}(\phi) = v_{2, \mu}(\phi),
\label{7.2.3}\end{equation}
for the bosonic states, and
\begin{equation}
\eta_\mu^{A}(\phi) = t(\phi)\, v_{2, \mu}(\phi),
\qquad
\eta_\mu^{B}(\phi) = s(\phi)\, v_{2, \mu}(\phi),
\label{7.2.4}\end{equation}
for the fermionic states.
As before, by using the relation
(\ref{7.1.6}) with $\nu$ substituted by $1-\nu$, one may replace $v_{2,
\mu}$
with the corresponding Gegenbauer polynomial
$C_l^{1-\nu}$ in the final expression of the
solutions.

\subsection{The case $U = \sigma_1$}

The case $U = \sigma_1$ is distinguished in the sense that it leads to the
free connection condition (i.e., both the wave function and its
derivative are continuous) in the limit $\nu \to 1$ where the singularity of
the potential disappears.
This `free' case is one of the non-separating cases in which
the spectrum consists of eigenvalues of both type 1 and type 2 eigenstates.
The eigenvalues are determined by the spectral condition (\ref{4.45}),
which now simplifies to
\begin{equation}
-\frac{\cos\pi\mu}{\cos\pi\nu}= \Re(\tau),
\label{7.1}\end{equation}
using that the parameters in (\ref{3.18}) are ${\cal A} = 0$ and
${\cal B} = 1$.

{}For type 1 eigenstates for which $\Re(\tau) = \pm 1$, it is immediate to
find the solutions for positive $\mu$.  For $\Re(\tau) = 1$, these are
\begin{equation}
\mu = 2n + 1 + \nu
\quad\hbox{and}\quad
\mu = \vert 2n + (1-\nu)\vert,
\qquad n = 0, 1, 2, \ldots,
\label{7.2}\end{equation}
which correspond to the $A_-$ and the $B_+$ case defined in (\ref{4.47}) and
(\ref{4.48}) (see also (\ref{4.24}), (\ref{4.29})).  Similarly,
for $\Re(\tau) = -1$, we obtain
\begin{equation}
\mu = 2n + \nu
\quad\hbox{and}\quad
\mu = 2n + 1 + (1-\nu),
\qquad n = 0, 1, 2, \ldots,
\label{7.3}\end{equation}
which correspond to the $B_-$ and the $A_+$ case.
In fact, these are the solutions mentioned in
(\ref{4.51})--(\ref{4.56}),
which arise since our parameters (\ref{3.18}) satisfy
$e^{i(\alpha + \beta)} = 1$ and $e^{i(\alpha - \beta)} = -1$.
Thus, the type 1 solutions for the angular part are
\begin{eqnarray}
&&
\eta_\mu^{A_+}(\phi) =
-c(\phi)\, a_1(\mu) v_{2, \mu}(\phi),
\qquad \mu = 2n + 1 + (1-\nu),
\nonumber\\
&&
\eta_\mu^{A_-}(\phi) =
t(\phi)\, a_2(\mu) v_{1, \mu}(\phi),
\qquad\quad\! \mu = 2n + 1 + \nu,
\nonumber\\
&&
\eta_\mu^{B_+}(\phi) =
-b_1(\mu) v_{2, \mu}(\phi),
\qquad\qquad \mu = \vert 2n + (1-\nu)\vert,
\nonumber\\
&&
\eta_\mu^{B_-}(\phi) =
s(\phi)\, b_2(\mu) v_{1, \mu}(\phi).
\qquad\quad \mu = 2n + \nu.
\label{7.4}\end{eqnarray}

Combining these with the eigenstates for the radial part, we obtain the type
1 eigenfunctions and energy eigenvalues for the entire system ($m, n = 0, 1,
2, \ldots$):
\begin{eqnarray}
&& \Psi^{++}_{mn}(r, \phi) = 
R_{m, \lambda}(r)\,\eta_\mu^{B_+}(\phi),
\qquad
E^{++}_{mn} = 2c\left(2m + 1 + 3\vert 2n + (1 - \nu) \vert\right),
\nonumber\\
&& \Psi^{-+}_{mn}(r, \phi) = 
R_{m, \lambda}(r)\,\eta_\mu^{A_+}(\phi),
\qquad
E^{-+}_{mn} = 2c\left(2m + 1 + 3( 2n + 1 + (1 - \nu)) \right),
\nonumber\\
&&
\Psi^{+-}_{mn}(r, \phi) = 
R_{m, \lambda}(r)\,\eta_\mu^{B_-}(\phi),
\qquad
E^{+-}_{mn} = 2c\left(2m + 1 + 3( 2n + \nu )\right),
\nonumber\\
&&
\Psi^{--}_{mn}(r, \phi) = 
R_{m, \lambda}(r)\,\eta_\mu^{A_-}(\phi),
\qquad
E^{--}_{mn} = 2c\left(2m + 1 + 3( 2n + 1 + \nu )\right),
\label{7.8}\end{eqnarray}
where the superscripts on $\Psi_{mn}$ specify the $D_6$ representation
similarly to (\ref{4.56.1}).
We observe that for $U=\sigma_1$ the type 1 eigenstates $\eta_\mu^{A_+}$,
$\eta_\mu^{B_+}$
are basically the bosonic eigenstates (\ref{7.2.3})
admitted under $U ={\bf 1}_2$, while $\eta_\mu^{A_-}$, $\eta_\mu^{B_-}$ are
the fermionic eigenstates
(\ref{7.1.5}) admitted under $U = -{\bf 1}_2$.

Next, we turn to type 2 eigenstates for which $\Re(\tau) = \pm {1\over 2}$.
Since the spectral condition (\ref{7.1}) is analogous to the type 1 case,
if we use
\begin{equation}
\Delta(\nu) := {1\over\pi}\arccos\left(\frac{1}{2}\cos\pi\nu\right),
\label{7.9}\end{equation}
we obtain the solutions,
\begin{equation}
\mu = 2n + 1 + \Delta(\nu)
\quad\hbox{and}\quad
\mu = 2n + (1-\Delta(\nu)),
\qquad n = 0, 1, 2, \ldots,
\label{7.10}\end{equation}
for $\Re(\tau) = {1\over 2}$, and
\begin{equation}
\mu = 2n + \Delta(\nu)
\quad\hbox{and}\quad
\mu = 2n + 1 + (1-\Delta(\nu)),
\qquad n = 0, 1, 2, \ldots,
\label{7.11}\end{equation}
for $\Re(\tau) = -{1\over 2}$.
Note that
\begin{equation}
{1\over 2} < \Delta(\nu) < {2\over 3}, \qquad \Delta(\nu) < \nu
\label{7.12}\end{equation}
for $\nu$ in the range (\ref{2.9}).
We also remark that no solution with $\mu^2 \leq 0$
is allowed for (\ref{7.1}) for both type 1 and type 2 eigenstates.

The eigenfunctions associated with the type 2 eigenvalues can be
constructed by
the procedure of Section 4.3.   Namely, one first obtains the eigenvector
(\ref{4.65.3})  with the aid of the projection operator
$\pi_\tau(\mu)$ in (\ref{4.65.4}).
Then, taking into account (\ref{4.31}) and (\ref{4.43}),
one forms the eigenfunction
$\eta_\mu(\phi)$ in (\ref{4.9})
out of the functions $\eta_{\pm, \mu}^k(\phi)$.
 Using the spectral
condition (\ref{7.1}) and $y(\mu) = -4\ri a_1(\mu)a_2(\mu)$,
$z(\mu) = -4\ri b_1(\mu)b_2(\mu)$ for $\pi_\tau(\mu)$,
and choosing the overall scale factor
of the eigenvector appropriately, one arrives at
\begin{equation}
\eta_{\mu,\tau}(\phi) = - {{\ri q(\mu)}\over{\Im(\tau)}} v_{1,\mu}(\phi)  +
v_{2,\mu}(\phi) ,
\label{7.12-1}\end{equation}
where
\begin{equation}
q(\mu) = {{3\cos^2\pi\nu}\over{2\pi^2}} 2^{-2\nu}\,
\Gamma(-\nu+{1\over 2}) \,\Gamma(-\nu+{3\over 2})\,\Gamma(\nu+
\mu)\,\Gamma(\nu- \mu).
\label{7.12-2}\end{equation}
This is valid for $0 < \phi \le {\pi\over 6}$ in sector 1, and extension to
the remaining half of sector 1 can be done by expressing in (\ref{7.12-1})
the
$v_{i,\mu}$ on $0< \phi \leq \frac{\pi}{6}$ in terms of the
functions $\eta_{\pm, \mu}^1(\phi)$ defined by (\ref{4.5}), (\ref{4.6}),
and adopting the resulting formula on the whole sector 1.
Subsequent
extension to sector $k$ can be made in terms of the rotated functions
$\eta_{\pm, \mu}^k(\phi)$ defined in (\ref{4.7}) and the eigenvector for
sector $k$ which is given by $\tau^{k-1}$ times the eigenvector for sector 1
(see (\ref{4.31}) and (\ref{4.43})).  Note that to each $\mu$ we have two
solutions for $\eta_{\mu,\tau}(\phi) $ on account of $\Im(\tau) = \pm
{{\sqrt{3}}\over 2}$, implying that
each  level is indeed doubly degenerate.

To display the above eigenstates and their
eigenvalues more systematically, let us use
the notation introduced in the paragraph above (\ref{4.62}) for the states
belonging to the different type 2 representations of $D_6$.
Thus the states $\eta^{(2)}_{\mu,\tau}$
arise for $\Re(\tau) = {1\over 2}$ and $\tilde
\eta^{(2)}_{\mu,\tau}$ for $\Re(\tau) = - {1\over 2}$,
that is,  for eigenvalues
(\ref{7.10}) and (\ref{7.11}), respectively.   We observe that, like in the
case of type 1 states,  each of the sets $\{ \eta^{(2)}_{\mu,\tau}\}$ and
$\{ \tilde \eta^{(2)}_{\mu,\tau}\}$ can be
classified into two distinct series according to the difference in the
non-integral part of $\mu$.   We introduce the notation
$\eta^{(2)-}_{\mu,\tau}$ for the eigenstates  (\ref{7.12-1})
with $\mu = 2n + 1 + \Delta(\nu)$ and $\eta^{(2)+}_{\mu,\tau}$ for
those with $\mu =  2n + (1-\Delta(\nu))$, and similarly
$\tilde \eta^{(2)-}_{\mu,\tau}$ with $\mu =
2n + \Delta(\nu)$ and $\tilde \eta^{(2)+}_{\mu,\tau}$ with $\mu = 2n + 1 +
(1-\Delta(\nu))$.
Combining with the solutions for the radial part, and adopting  similar
notation to specify the entire eigenstates containing  the
type 2 angular states, we
obtain
\begin{eqnarray}
&& \Psi^{(2)+}_{mn, \tau}(r, \phi) =
R_{m, \lambda}(r)\,\eta_{\mu,\tau}^{(2)+}(\phi),
\qquad
E^{(2)+}_{mn} = 2c\left(2m + 1 + 3( 2n + (1-\Delta(\nu)))\right),
\nonumber\\
&& \tilde \Psi^{(2)+}_{mn,\tau}(r, \phi) =
R_{m, \lambda}(r)\,\tilde \eta_{\mu,\tau}^{(2)+}(\phi),
\qquad
\tilde E^{(2)+}_{mn} = 2c\left(2m + 1 + 3(2n + 1 + (1-\Delta(\nu))\right),
\nonumber\\
&& \tilde \Psi^{(2)-}_{mn,\tau}(r, \phi) =
R_{m, \lambda}(r)\,\tilde \eta_{\mu,\tau}^{(2)-}(\phi),
\qquad
\tilde E^{(2)-}_{mn} = 2c\left(2m + 1 + 3(2n + \Delta(\nu))\right),
\nonumber\\
&& \Psi^{(2)-}_{mn,\tau}(r, \phi) =
R_{m, \lambda}(r)\,\eta_{\mu,\tau}^{(2)-}(\phi),
\qquad
E^{(2)-}_{mn} = 2c\left(2m + 1 + 3(2n + 1 + \Delta(\nu))\right).
\label{7.13}\end{eqnarray}
Note that $\mu$ can be recovered from the energy  as
$3\mu= \frac{E}{2c} - 2m-1$.
The energy eigenvalues in (\ref{7.8}) and (\ref{7.13}) provide the complete
spectrum of the $N = 3$ Calogero model defined
by the Hamiltonian ${\hat H}_{rel}$ (\ref{2.3})
under the angular boundary condition $U = \sigma_1$.
We mention that, for any
$\nu$, the ground (the lowest energy) state is given by the type 1 state
$\Psi^{++}_{00}$ possessing the energy
$E^{++}_{00} = 2c\left(1 + 3\vert 1 - \nu\vert\right)$.

Now, let us consider the harmonic oscillator limit $\nu \to 1$.  Here,
the functions in (\ref{4.1}) reduce to
\begin{equation}
v_{1,\mu}(\phi) = {1\over \mu}  \sin{3\mu\phi},
\qquad
v_{2,\mu}(\phi) =  \cos{3\mu\phi} ,
\label{7.14}\end{equation}
for sector 1, and we have
\begin{eqnarray}
&&
a_1(\mu) = {1\over \mu} \sin{\left({{\pi\mu}\over 2}\right)},
\qquad
a_2(\mu) = \cos{\left({{\pi\mu}\over 2}\right)},
\nonumber\\
&&
b_1(\mu) = 3 \cos{\left({{\pi\mu}\over 2}\right)},
\qquad\,\,
b_2(\mu) = -3\mu \sin{\left({{\pi\mu}\over 2}\right)}.
\label{7.15}\end{eqnarray}
These are either zero or proportional to $(-1)^n$
for $\mu$ in (\ref{7.4}) with $\nu=1$,
and hence
the type 1 states are basically given by the trigonometric functions
(\ref{7.14}).
Notice that in the $\nu=1$ limit,
except for $n = 0$, the states $\eta_\mu^{B_+}$ and
$\eta_\mu^{A_-}$ are degenerate
with eigenvalue $\mu = 2n$, and similarly $\eta_\mu^{A_+}$ and
$\eta_\mu^{B_-}$ are degenerate with $\mu = 2n + 1$.  These two pairs of
degenerate states also share the same eigenvalue among themselves for
$\hat{\cal T}$ with $\tau = 1$ and $\tau = -1$, respectively. Thus, one may
form their linear combination to obtain the simpler set of eigenstates
$e^{\pm \ri 3\mu \phi}$ for integers  $\mu = 1, 2, \ldots$.  These states have
$\tau = 1$ for $\mu$ even and $\tau = -1$ for $\mu$ odd.

To find the type 2 states in the limit,  we observe that
$\Delta(\nu) \to {2\over 3}$ for  $\nu \to 1$, and that the factor $q(\mu)$
in (\ref{7.12-2}) reduces  to
\begin{equation}
q(\mu) = -{3\over 4}{{\mu}\over{\sin\pi\mu}}.
\label{7.16.1}\end{equation}
Hence, for the eigenvalues $\mu$ in  (\ref{7.10}),  (\ref{7.11}), the
solution
(\ref{7.12-1}) becomes
$\eta_{\mu,\tau}(\phi) = e^{\pm \ri 3\mu \phi}$,
which is valid for all sectors,
where
the signs $\pm $ correspond to $\Im(\tau) = \pm {{\sqrt{3}}\over 2}$ for
$\eta_{\mu,\tau} = \eta_{\mu,\tau}^{(2)+}$ and
$\tilde \eta_{\mu,\tau}^{(2)-} $ and to $\Im(\tau)
= \mp {{\sqrt{3}}\over 2}$ for
$\eta_{\mu,\tau} = \eta_{\mu,\tau}^{(2)-}$ and
$\tilde \eta_{\mu,\tau}^{(2)+} $.

Consequently, if we introduce $k := 3\mu$ (which yield integers for all $\mu$
as $\nu \to 1$),
for $\nu=1$ both the type 1 and type 2 eigenstates can be
combined to be presented together as
\begin{equation}
\eta_k^\pm(\phi) := e^{\pm \ri k \phi},
\qquad
k = 0, 1, 2, \ldots.
\label{7.18}\end{equation}
In the $\nu=1$ limit, the complete set of  eigenfunctions
and eigenvalues of $\hat H_{rel}$ (\ref{2.3}) is therefore furnished by
\begin{equation}
\Psi^\pm_{mk}(r, \phi) =
R_{m, \lambda}(r)\,\eta_k^\pm(\phi),
\qquad
E^\pm_{mk} = 2c\left(2m + 1 + k \right),
\label{7.19}\end{equation}
for $m$, $k = 0, 1, 2, \ldots$, where it understood that both signs give the
same if $k=0$.
 In view of $\lambda = k^2$, we see that
\begin{equation}
R_{m, \lambda}(r) = r^{k} e^{-\frac{1}{2} c  r^2} L_m^{k}(c r^2).
\label{7.20}\end{equation}
The states (\ref{7.18}) are the $2\pi$-periodic
eigenstates of the operator $M$ in (\ref{2.6}) for $g =0$ without
singularity at the points of ${\cal S}$ (\ref{3.5}).
Correspondingly, the
eigenfunctions and the eigenvalues  (\ref{7.19}) recover  precisely
the ones known for
the harmonic oscillator in 2-dimensions
(see, e.g., \cite{Levin} for comparison).  This
shows that
under the `free' boundary condition $U = \sigma_1$ our system is smoothly
connected to the harmonic oscillator in the limit $\nu \to 1$.
This is not the case for the previous two
cases, $U = -{\bf 1}_2$ and $U = {\bf 1}_2$.
Indeed, these become such systems in the  $\nu \to 1$
limit in which  `two thirds' of the levels of the harmonic oscillator
are missing and each level has multiplicity 6
(instead of 2) in the angular sector.

\subsection{The case $U = -\sigma_1$}

\bigskip
\begin{figure}  \centering  
\resizebox{0.99\textwidth}{!}{\includegraphics{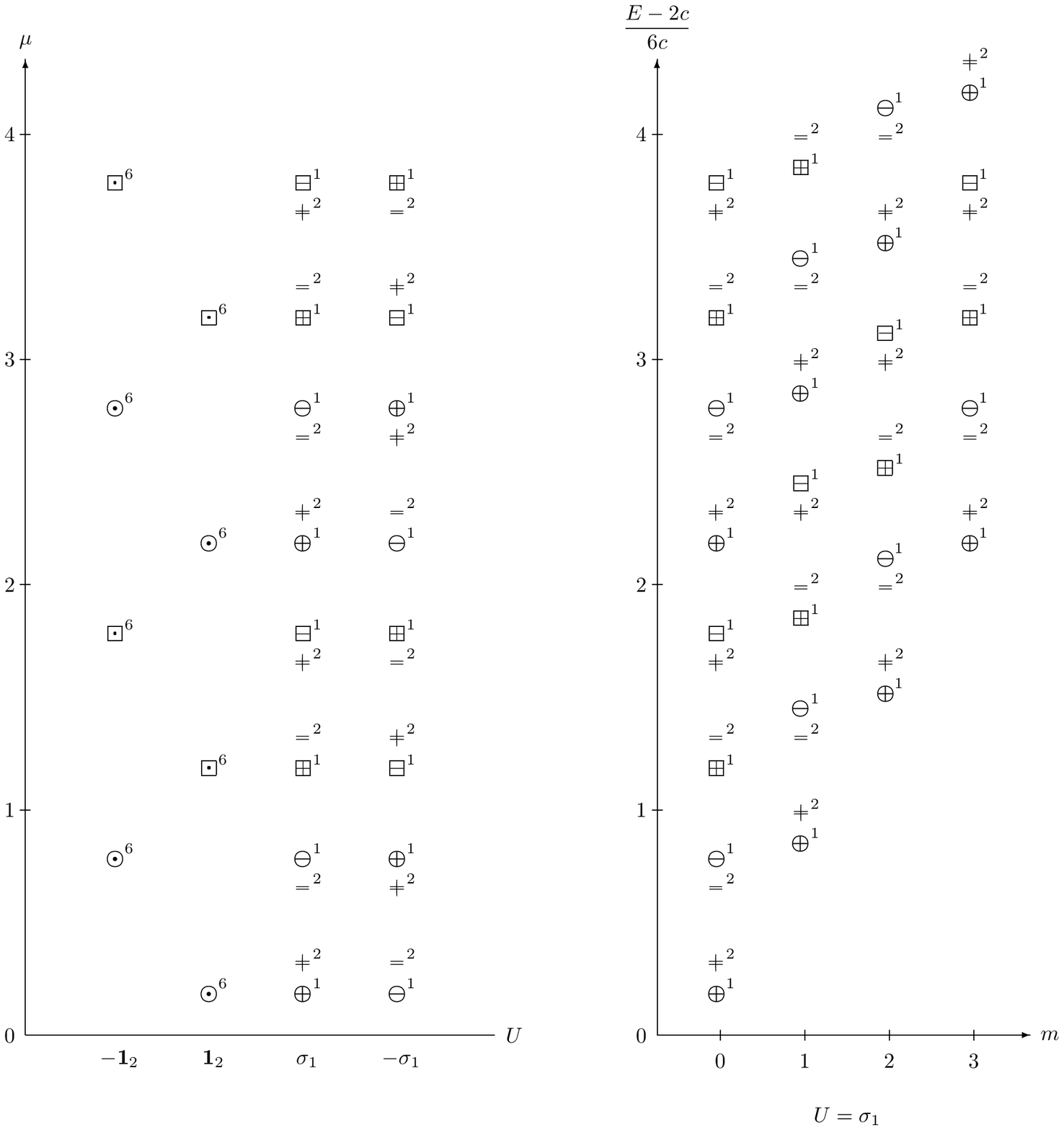}}
\caption{
Left: The angular `eigenvalue-parameter' $\mu$
for the four explicitly solvable cases discussed, with $\nu=4/5$.
Boxes stand for `case A'  states and circles for
`case B' ones; the $+$ or $-$ within them shows the sign of $\Re(\tau)$, a dot
is applied when the sign is undetermined.
The symbols
$\setbox1\hbox{$+$}\copy1\hskip-\wd1\raise .3ex\box1$
and
$\setbox1\hbox{$-$}\copy1\hskip-\wd1\raise .3ex\box1$
indicate type
2 states with positive and, respectively, negative $\Re(\tau)$. The
superscript numbers display the multiplicities. Right: the energy
spectrum (conveniently shifted and rescaled) for the case $U = \sigma_1$.
}
\label{fig:spectral}  \end{figure}
\bigskip

The case $U = -\sigma_1$,
where we have ${\cal A} = 0$ and ${\cal B} = -1$ in (\ref{3.18}),
can be dealt with analogously to the `free' case
$U = \sigma_1$.  Indeed,
the spectral condition (\ref{4.45}) now reads
\begin{equation}
\frac{\cos\pi\mu}{\cos\pi\nu}= \Re(\tau),
\label{7.21}\end{equation}
and hence, as a whole, the spectrum remains the same as that of the `free'
case.  The only difference is that, because of the opposite sign in
(\ref{7.21}) on the right hand side compared to (\ref{7.1}), the eigenvalues
associated with the solutions are interchanged.
For type 1 eigenstates, the interchange amounts to
$A_\pm \to A_\mp$ and $B_\pm \to B_\mp$.
Thus, the angular solutions become
\begin{eqnarray}
\eta_\mu^{A_+}(\phi) =
c(\phi)\, a_2(\mu) v_{1, \mu}(\phi),
\qquad\,
&&\mu = 2n + 1 + \nu,
\nonumber\\
\eta_\mu^{A_-}(\phi) =
-t(\phi)\, a_1(\mu) v_{2, \mu}(\phi),
\quad \,\,\,
&&\mu = 2n + 1 + (1-\nu),
\nonumber\\
\eta_\mu^{B_+}(\phi) =
b_2(\mu) v_{1, \mu}(\phi),
\qquad\qquad \,\,
&&\mu = 2n + \nu,
\nonumber\\
\eta_\mu^{B_-}(\phi) =
-s(\phi)\, b_1(\mu) v_{2, \mu}(\phi),
\quad \,\,\,
&&\mu = \vert 2n + (1-\nu)\vert.
\label{7.22}\end{eqnarray}
Accordingly, the type 1 eigenfunctions for the entire system are obtained
from
(\ref{7.8}) with the interchange of eigenstates and eigenvalues $\mu$
as shown in (\ref{7.22}).

The type 2 eigenstates acquire a similar change as observed for type 1
states.  Explicitly, the solutions for the spectral condition are given by
(\ref{7.10}) and (\ref{7.11}) with the interchange of the cases
$\Re(\tau) = {1\over 2}$ and $\Re(\tau) = - {1\over 2}$, i.e., the states
$\eta^{(2)}$ and
$\tilde \eta^{(2)}$ are swapped.  Consequently, the type 2 eigenfunctions
of the entire system are obtained from the solutions for the case
$U = \sigma_1$ by the corresponding interchange of eigenstates and
eigenvalues $\mu$.

Finally, we mention that if  $U
= - \sigma_1$, then the system does not tend to
the 2-dimensional harmonic oscillator as $\nu \to 1$, even
though the spectrum reduces
to that of the harmonic oscillator in this limit.
 This can be seen, for instance, by
looking at the ground state wave function,
$R_{0,\lambda}(r)\,\eta_\mu^{B_-}(\phi)$ with $\mu=\vert 1-\nu\vert$,
which has parity $-1$
under  $\hat P_n$ for any $n$ in disagreement
with  the $+1$ parity of the oscillator ground state.

\section{Conclusion}
\setcounter{equation}{0}

In this paper we explored the inequivalent
quantizations of the three-particle Calogero model in the separation
of variables approach under the assumption (\ref{2.9}) on the coupling
constant.
Upon requiring the $D_6$ symmetry, we found that the model permits
inequivalent quantizations for the angular Hamiltonian $M$ (\ref{2.6})
which are specified by boundary conditions of the form (\ref{3.14})
parametrized by a matrix $U\in U(2)$ satisfying (\ref{3.15}).
We showed that the angular boundary conditions
fall into the qualitatively different `separating' and `non-separating'
classes, and it is possible only in the separating
case to set the admissible
wave functions to zero in all but one of the six sectors corresponding
to the different orderings of the particles.
Another important distinction was uncovered between the boundary
conditions admitting and the ones not admitting a negative
eigenvalue of $M^U$,
since in the former case the
energy is not bounded from below, which is in general not permissible
in physical applications.
The properties that $M^U$ has an eigenvalue $\lambda <0$
or that  it possesses only eigenvalues $\lambda >0$  are
stable generically (in the sense of Section 5.4)
with respect to small perturbations of the parameters of the `connection
matrix' $U$ (\ref{3.18}).
Our description of the inequivalent
quantizations of the
radial Hamiltonian (\ref{6.2}) for $0<\lambda <1$ is consistent with
and complements the previous analysis \cite{indPLA}.

We classified the eigenstates of the Hamiltonian
according to the irreducible representations of the
$D_6$ symmetry group, and described also
the induced classification under
the exchange-$S_3$ subgroup of $D_6$.
If necessary in some application, one can consistently
truncate the Hilbert space to a sector containing only the states of
a fixed symmetry type.
Our construction provides new quantizations also for the so-obtained
truncated sectors, containing for example the
states of `bosonic' or `fermionic' character
with respect to the permutations of the particles.

Our case-study illustrates the fact that
inequivalent quantizations have very different properties in general,
and external theoretical or experimental input is needed
to choose between such quantizations.
One possible criterion for the choice may be the smoothness of the model in the
limit where the singularity of the potential disappears.
Our solution for $U = \sigma_1$ mentioned in Section 7 shows that
there indeed exists a distinguished quantization that meets this
criterion.

Of course, the present work can only be regarded as a
`theoretical laboratory'
since most applications of the Calogero model use
arbitrarily large particle number.
It would be very interesting to extend our construction to
the $N$ particle case, which would require understanding
the possible self-adjoint domains of the partial differential
operator $H_\Omega$ in (\ref{1.6}) under
the assumption (\ref{1.3}).
For example, we wonder if an analogue of the explicitly solvable
`free' case that we found for $N=3$ exists for general $N$.
It would be also interesting to better understand
the inequivalent self-adjoint domains of the Hamiltonian
without adopting the separation of variables approach,
starting directly from the minimal operator
corresponding to the formal expression (\ref{1.1}).

\bigskip
\bigskip
\noindent{\bf Note added.}
We learned after submitting the paper that the spectra of the
self-adjoint extensions of $\cH_{r,\lambda}$ (\ref{6.2}) have
also been studied, for $\frac{1}{4}\leq \lambda <1$,
in \cite{FPW}.  The method used in \cite{FPW}  is similar to
that used in \cite{indPLA}, and the results are consistent with
our results derived in Section 6 relying on a different,
but equivalent, method.  We thank P.A.G.  Pisani for drawing our
attention to this article.

\bigskip
\bigskip
\noindent{\bf Acknowledgements.}
This work was supported in part by the Hungarian
Scientific Research Fund (OTKA) under grant numbers
T034170, T043159, T049495, M36803, M045596  and by the EC network `EUCLID',
contract number HPRN-CT-2002-00325.
It was also supported by the Grant-in-Aid for Scientific Research,
No.~13135206 and No.~16540354, of the Japanese Ministry of Education,
Science, Sports and Culture.

\renewcommand{\theequation}{\arabic{section}.\arabic{equation}}
\renewcommand{\thesection}{\Alph{section}}
\setcounter{section}{0}
\section{Remarks on the angular and radial Hamiltonians}
\renewcommand{\theequation}{A.\arabic{equation}}

The characterization of the self-adjoint domains for the angular
Hamiltonian described in Section 3  can be viewed as
an application of the general theory of self-adjoint
differential operators \cite{DS,Kochubei,Gorba}.
Nevertheless, it  may be useful to present
an elementary argument proving that the conditions in (\ref{3.14})
provide self-adjoint domains for $M$ (\ref{2.6}).
In this appendix we also wish to quote some theorems from \cite{DS}
that imply the discreteness of the spectrum
of the radial Hamiltonian $\cH_{r,\lambda}$ (\ref{6.2})
on any self-adjoint domain.

It was mentioned in Section~3 that
the self-adjoint domains for $M$ arise as restrictions
of the maximal domain $\cD_1$.
Here, our aim is to show that the restriction
$\cD \subset \cD_1$ defined by the conditions in (\ref{3.14}) yields
a self-adjoint domain, i.e.,  the restriction of $M_{\cD_1}$ to $\cD$
is a self-adjoint operator.
For this, it proves advantageous to rewrite (\ref{3.14}) in
the equivalent form
  \begin{equation}
  U_{\x} B^{(+)}_{\x}(\psi) =
  B^{(-)}_{\x}(\psi) \quad (U_{\x}\in U(2),\,  {\x} \in \cS)\,
  \quad \hbox{with} \quad
  B^{(\pm)}_{\x}(\psi) := B_{\x}(\psi) \pm \ri B'_{\x}(\psi) \, .
  \label{A.3}\end{equation}
Note that all the twelve `boundary vectors' $B^{(\pm)}_{\x}(\psi)$ take
independently all the possible ${\bf C}^2$ vector values as $\psi$ runs
over $\cD_1$.  One can see this by  considering
the boundary vectors associated with the functions
  \be
  [ c^{}_1 \varphi_1^{\x} + c^{}_2 \varphi_2^{\x} ] \eta,
  \label{A.4}\ee
where $c_1, c_2 \in {\bf C}$, ${\x}\in \cS$ (\ref{3.5})
is one of the singular points,
and $\eta \in C^\infty (S^1 \setminus \cS)$ is a function
taking the constant value $1$ on one side of ${\x}$ on a small closed
interval and being zero on the other side of ${\x}$ as well as on both
sides of the five other singular points.

Next, let us point out that, for $\forall \psi, \eta \in \cD_1$,
  \bea
  ( \psi, M_{\cD_1} \eta ) - ( M_{\cD_1} \psi, \eta )
  & = & \sum\limits_{{\x} \in \cS}
  \left( W[ \bar \psi, \eta ]_{{\x}+} - W[ \bar\psi, \eta ]_{{\x}-} \right) \nn
  & = & \sum\limits_{{\x} \in \cS}
  \left( \langle B_{\x}(\psi), B'_{\x}(\eta) \rangle
  - \langle B_{\x}(\psi), B'_{\x}(\eta) \rangle  \right) \label{A.5} \\
  & = & \frac{1}{2\ri} \sum\limits_{{\x} \in \cS}
  \left( \langle B^{(+)}_{\x}(\psi), B^{(+)}_{\x}(\eta) \rangle
  - \langle B^{(-)}_{\x}(\psi), B^{(-)}_{\x}(\eta) \rangle \right),
  \nonumber  \eea
where $(\cdot,\cdot)$ is the scalar product in $L^2(S^1)$ and
$\langle\cdot,\cdot\rangle$ is the scalar product in ${\bf C}^2$.
Formula (\ref{A.5})
can be derived by partial integration using the identity
  \be
  W[ \bar \psi, \eta ] = W[ \bar\psi, \varphi_1^{\x} ] W[ \eta, \varphi_2^{\x} ] -
  W[ \bar\psi, \varphi_2^{\x} ] W[ \eta, \varphi_1^{\x} ] \, ,
  \label{A.6}  \ee
which is valid on the domain of definition of the reference modes
$\varphi_k^{\x}$  as a result of
$W[ \varphi_1^{\x}, \varphi_2^{\x} ] = 1$ (see Section~3).
It  follows from (\ref{A.3}) that, for $\psi, \eta \in \cD$,
the expression given by (\ref{A.5}) vanishes
(in fact, each term in the sum vanishes
separately). This means that $\cD$ is a symmetric domain within
$\cD_1$, i.e., $M_\cD$ is a symmetric operator.
To
demonstrate that this domain is a self-adjoint one,
it is enough to show
that the vanishing of (\ref{A.5}) for all $\psi \in \cD$ with a fixed
$\eta\in \cD_1$ implies that $\eta\in \cD$.

We now choose two functions $\psi_1, \psi_2 \in \cD$ such that, for a given
${\x} \in \cS$, $ B^{(+)}_{\x}(\psi_1) $ and $ B^{(+)}_{\x}(\psi_2) $ form an
orthonormal basis in ${\bf C}^2$ [by (\ref{A.3}), $ B^{(-)}_{\x}(\psi_1) $
and $ B^{(-)}_{\x}(\psi_2) $ then also form an orthonormal basis] and
the boundary vectors at the other singular points are zero. If
  \be
  ( \psi_k, M_{\cD_1} \eta ) - ( M_{\cD_1} \psi_k, \eta )
  = \frac{1}{2\ri}
  [ \langle B^{(+)}_{\x}(\psi_k), B^{(+)}_{\x}(\eta) \rangle
  - \langle B^{(-)}_{\x}(\psi_k), B^{(-)}_{\x}(\eta) \rangle ]
  \label{A.7}  \ee
is zero for $k = 1, 2$, then we can write
  \bea
  U_{\x} B^{(+)}_{\x}(\eta) & = & U_{\x} \left[
  \langle B^{(+)}_{\x}(\psi_1), B^{(+)}_{\x}(\eta) \rangle
  B^{(+)}_{\x}(\psi_1) +
  \langle B^{(+)}_{\x}(\psi_2), B^{(+)}_{\x}(\eta) \rangle
  B^{(+)}_{\x}(\psi_2) \right] \nn & = &
  \langle B^{(+)}_{\x}(\psi_1), B^{(+)}_{\x}(\eta) \rangle
  U_{\x} B^{(+)}_{\x}(\psi_1) +
  \langle B^{(+)}_{\x}(\psi_2), B^{(+)}_{\x}(\eta) \rangle
  U_{\x} B^{(+)}_{\x}(\psi_2) \nn  & = &
  \langle B^{(-)}_{\x}(\psi_1), B^{(-)}_{\x}(\eta) \rangle
  B^{(-)}_{\x}(\psi_1) +
  \langle B^{(-)}_{\x}(\psi_2), B^{(-)}_{\x}(\eta) \rangle
  B^{(-)}_{\x}(\psi_2) \nn & = & B^{(-)}_{\x}(\eta).
  \label{A.8}  \eea
This implies  that $\eta\in \cD$ as required.

The self-adjoint domains  for the formal radial Hamiltonian
$\cH_{r,\lambda}$ (\ref{6.2})
can be treated  similarly to the above,
and this case is actually much simpler.
Since the boundary condition (\ref{6.6}) appears in several references
\cite{Richt,Krall,Kochubei}, we need not dwell on this point.
We below summarize the general results that imply the discreteness
of the spectrum of the radial Hamiltonian on any of these
self-adjoint domains.

Recall that the `discrete spectrum' of a self-adjoint operator
consists of the isolated points of the spectrum that are
eigenvalues of finite multiplicity,
and the rest of the spectrum is called the `essential spectrum'.
(Note that the isolated points of the spectrum are always
eigenvalues, and for ordinary differential operators all
eigenvalues have finite multiplicity.)
In the case of self-adjoint ordinary differential operators
the essential spectrum is the same for all
self-adjoint extensions  of the `minimal operator',
and thus it can be assigned unambiguously to the underlying
formal differential operator (see e.g. XIII.6.4 in \cite{DS}).
According to the statement of XIII.7.4 \cite{DS},
the essential spectrum of the formal differential operator
$\cH_{r,\lambda}$  on the interval $(0,\infty)$ decomposes as the union of the
essential spectra of the operators of the same form on
$(0, x_0]$ and on $[x_0, \infty)$ for any  $x_0>0$.
The essential spectrum assigned to the interval $[x_0,\infty)$ is
empty by XIII.7.16 \cite{DS}, since the potential term of
$\cH_{r,\lambda}$ tends to $+\infty$ as $r\to \infty$.
If $\lambda > \frac{1}{4}$, then the potential also tends to $+\infty$
as $r\to 0$, and the essential spectrum associated to
$(0,x_0]$  is therefore empty by XIII.7.17 \cite{DS}.
If $\lambda <1$, the same conclusion follows from XIII.6.12 in \cite{DS}
by using that the deficiency indices of
the minimal operator on $(0,x_0]$ are $(2,2)$.
By combining these, we see that the essential spectrum of the formal
differential operator
 $\cH_{r,\lambda}$ on $(0, \infty)$ is empty,
and hence all of its self-adjoint versions have pure discrete spectra.

The above arguments can be used to prove the
discreteness of the spectrum of the radial Hamiltonian for any particle
number $N$, since $H_{r,\lambda}$ given by (\ref{1.5}),
(\ref{1.7}) leads to the equivalent operator in $L^2({\bf R}_+, dr)$
(see eq.~(\ref{6.2}))
\begin{equation}
\cH_{r,\lambda} := r^{\frac{N-2}{2}} \circ H_{r,\lambda}
\circ r^{\frac{2-N}{2}} =
- \frac{\hbar^2}{2m}\frac{d^2}{d   r^2} +
\frac{N}{4} m\omega^2 r^2 +\frac{\hbar^2}{8m}(N-2)(N-4)/r^2 + \lambda/r^2,
\label{A.9}\end{equation}
which must be a self-adjoint
operator in  $L^2({\bf R}_+, dr)$.
For $N=3$ the discreteness of the spectrum of the angular Hamiltonian
also follows by similar reasoning,  but for $N>3$
 $H_\Omega$ in (\ref{1.6})
becomes a partial differential operator that would
require a different treatment.

\section{Representations of the symmetry group $D_6$}
\setcounter{equation}{0}
\renewcommand{\theequation}{B.\arabic{equation}}

The dihedral group $D_6$ admits four different 1-dimensional
representations and two inequivalent 2-dimensional
irreducible representations.
This follows since the 12 elements of $D_6$ fall into 6 conjugacy
classes as  described in  Figure \ref{fig:D6}  (with the notations in
eqs.~(\ref{3.1})--(\ref{3.3})),
and $12= 1 + 1 + 1 +1 + 2^2 + 2^2$.


\begin{figure}[ht]  \centering  
\renewcommand{\arraystretch}{1.4}
{\begin{tabular}{|ccccccc|}  \hline
$ \hbox{conjugacy} \atop \hbox{class} $\rule[-1.9ex]{0ex}{5.4ex} &
$\{e\}$ & $\{R_i\}$ & $\{P_i\}$ &
$\{{\cal R}_{\pi \hskip-.2ex / \hskip-.1ex 3}^{\pm 1}\}$ &
$\{{\cal R}_{\pi \hskip-.2ex / \hskip-.1ex 3}^{\pm 2}\}$ &
$\{{\cal R}_{\pi \hskip-.2ex / \hskip-.1ex 3}^3 \}$\\ \hline
$\chi^{++}$      &
\hbox{\null\hskip 12.85pt1\hskip 12.85pt\null}  &
\hbox{\null\hskip 12.85pt1\hskip 12.85pt\null}  &
\hbox{\null\hskip 12.85pt1\hskip 12.85pt\null}  &
\hbox{\null\hskip 12.85pt1\hskip 12.85pt\null}  &
\hbox{\null\hskip 12.85pt1\hskip 12.85pt\null}  &
\hbox{\null\hskip 12.85pt1\hskip 12.85pt\null}\\ \hline
$\chi^{-+}$      &   1  &   -1  &  1  &  -1  &   1  &  -1\\ \hline
$\chi^{+-}$      &   1  &  1  &   -1  &  -1  &   1  &  -1\\ \hline
$\chi^{--}$      &   1  &  -1  &  -1  &   1  &   1  &   1\\ \hline
$\chi^{(2)}$          &   2  &   0  &   0  &   1  &  -1  &  -2\\ \hline
$\tilde{\chi}^{(2)}$  &   2  &   0  &   0  &  -1  &  -1  &   2\\ \hline
\end{tabular}}  \caption{Character table of the group $D_6$.}
\label{fig:D6}  \end{figure}

\medskip

The $1$-dimensional (or `type 1') representation of character
$\chi^{\varrho p}$ with $\varrho,p\in \{\pm\}$
is defined by assigning the  parities
$\varrho$ and $p$ to the
reflections $R_k$ and $P_k$ ($k=1,2,3$), respectively.
Since the $P_k$ generate the exchange-$S_3$ subgroup of $D_6$,
the representations with $p=+$ can be called `bosonic' and
those with $p=-$ can be called `fermionic'.
The character of the $2$-dimensional defining representation of
$D_6$ is denoted by $\chi^{(2)}$.
The other $2$-dimensional (or `type 2') representation is the
tensor product of the defining representation and one of the
type 1 representations
with character $\chi^{+-}$ or $\chi^{-+}$.
The type 2 representations of $D_6$ remain irreducible
(and become equivalent) when restricted to the $S_3$ subgroups.

For reference in the main text, note that the eigenvalues
of $\cR^{\pm 1}_{\pi/3}$ in the defining representation are
$-\jmath$ and $-\bar\jmath$ and in the other type 2 representation
are $\jmath$ and $\bar\jmath$, with $\jmath = e^{\frac{2\pi \ri}{3}}$.
Indeed, this is a consequence of the relations
$\chi^{(2)}( \cR_{\pi/3}) = 1 = -\jmath - \bar \jmath$ and
$\tilde\chi^{(2)}( \cR_{\pi/3}) = -1 = \jmath +\bar \jmath$
taking into account that the eigenvalues of $\cR^{\pm 1}_{\pi/3}$
must be sixth roots of unity.

\section{The monotonicity of the  function $F_A$}
\setcounter{equation}{0}
\renewcommand{\theequation}{C.\arabic{equation}}

We here demonstrate that the function
$F_A$  defined in (\ref{5.2}) is strictly monotonically decreasing for
 $\mu \in (\mu_m^\infty, \mu_{m+1}^\infty)$, with any
$m\geq 0$ in (\ref{5.14}),
as well as for   $\mu \in [0, \mu_0^\infty)$.

Consider the logarithmic derivative of $F_A$,
\begin{equation}
2\frac{F_A'(\mu ) }{F_A(\mu )}=
\psi\left(\frac{\nu +1+ \mu}{2}\right)
-\psi\left(\frac{\nu +1- \mu}{2}\right)
+\psi\left(\frac{2-\nu - \mu}{2}\right)
-\psi\left(\frac{2-\nu + \mu}{2}\right).
\label{D.1}\end{equation}
Remember that
\begin{equation}
\psi(1-z)= \psi(z) + \pi \cot \pi z,
\label{D.2}\end{equation}
where $\psi(z)$ is strictly monotonically increasing on the
positive real semi-axis,
$\cot\pi z$ is decreasing between  two consecutive singularities.
We can rewrite (\ref{D.1}) as
 \begin{eqnarray}
2\frac{F_A'(\mu) }{F_A(\mu)} &=
& \left[\psi\left(\frac{\nu +1+ \mu}{2}\right)
-\psi\left(\frac{2-\nu+ \mu}{2}\right)\right]\nonumber\\
\quad &+&
\left[\psi\left(\frac{\nu + \mu}{2}\right)
-\psi\left(\frac{1-\nu + \mu}{2}\right)\right]\nonumber\\
\quad &+&
\left[\pi \cot \frac{\pi}{2} (\nu + \mu)
-\pi \cot\frac{\pi}{2} (1-\nu + \mu)\right].
\label{D.3}\end{eqnarray}
Using that $ \frac{1}{2} < \nu < \frac{3}{2}$ (\ref{2.9}),
one sees that if
\begin{equation}
\mu > (2-\nu)= \mu_0^0,
\label{D.4}\end{equation}
then the arguments of the four $\psi$ functions in (\ref{D.3})
as well as the contributions of the first two lines of this
formula are positive.

Referring to (\ref{5.14}), (\ref{5.15}) for the notations, suppose now that
\begin{equation}
\mu_m^0 <\mu< \mu_{m}^\infty,
\label{D.5}\end{equation}
where the function $F_A$ is negative.
By using the periodicity of $\cot$,
we find that
\begin{equation}
 \cot\frac{\pi}{2}(\nu + \mu)
- \cot\frac{\pi}{2}(1-\nu + \mu) =
\cot \frac{\pi}{2}(\nu + \mu -2 m-2)
- \cot \frac{\pi}{2}(1-\nu + \mu -2 m)>0,
\label{D.6}\end{equation}
since for this range of $\mu$
\begin{equation}
0<(\nu + \mu - 2m-2)< (1-\nu + \mu-2 m) <2,
\label{D.7}\end{equation}
thanks to $\nu < \frac{3}{2}$ (\ref{2.9}).
This proves that $F'_A(\mu)<0$, whenever $F_A(\mu)<0$ (with $\mu >0$).

For the `positive branch'  $F_A(\mu)>0$, supposing that
\begin{equation}
\mu_m^\infty <\mu< \mu_{m+1}^0,
\label{D.8}\end{equation}
we obtain
\begin{equation}
 \cot \frac{\pi}{2}(\nu + \mu)
- \cot\frac{\pi}{2}(1-\nu + \mu)=
\cot \pi\gamma - \cot \pi \vartheta
\label{D.9}\end{equation}
with
\begin{equation}
\gamma:  =
 \frac{\nu + \mu}{2} - (m+1),
 \qquad
 \vartheta := \frac{1-\nu + \mu}{2}- (m+1).
\label{D.10}\end{equation}
The difference (\ref{D.9}) is negative, since
\begin{equation}
0< \vartheta < \gamma < 1,
\label{D.11}\end{equation}
thanks to (\ref{D.8}) and  $\nu > \frac{1}{2}$ (\ref{2.9}).
In order to combine the terms of different signs in (\ref{D.3}),
we may use the
following standard integral formulae:
\begin{equation}
\psi(p) - \psi(q)= \int_0^1  \frac{X^{q-1} - X^{p-1}}{1-X} dX,
\qquad
p,q>0,
\label{D.12}\end{equation}
\begin{equation}
\pi  \cot \pi q= \int_0^1  \frac{X^{q-1} - X^{-q}}{1-X} dX,
\qquad
0<q<1.
\label{D.13}\end{equation}
For $\mu$ in (\ref{D.8}),
putting these into (\ref{D.3}) using (\ref{D.9}), (\ref{D.11})
and that now $(1-\nu + \mu)>0$,
we get
\begin{equation}
2\frac{F_A'(\mu) }{F_A(\mu)} =\int_0^1  \frac{P(X)}{1-X} dX
\label{D.14}\end{equation}
with
\begin{equation}
P(X)= (1 - X^{m+1}) (X^\gamma - X^\vartheta)X^{-1} +
(1- X^{\gamma + \vartheta + m +\frac{1}{2}})
(X^{-\vartheta} - X^{-\gamma}),
\label{D.15}\end{equation}
as one can verify straightforwardly.
Because of (\ref{D.11}),  $P(X)<0$ for $0<X<1$.
This proves that $F_A'(\mu)<0$ for $\mu$ in (\ref{D.8}).

The only case left to consider is
\begin{equation}
0 <\mu< \mu_{0}^0=(2-\nu).
\label{D.16}\end{equation}
In this case the arguments of all four $\psi$ functions in
(\ref{D.1}) are positive,
and we may proceed  with the aid of an integral formula relying
on (\ref{D.12}).
Similar to (\ref{D.14}), we now obtain
\begin{equation}
P(X)= (X^\mu -1) ( 1 - X^{\nu -\frac{1}{2}}) X^{\frac{2-\nu -\mu}{2}} <0,
\label{D.17}\end{equation}
which completes the proof of our claim concerning
the strictly decreasing nature of $F_A$.


\end{document}